\documentclass{article} 
\usepackage{iclr2027_conference,times}

\usepackage{amsmath,amsfonts,bm}

\def\eqref#1{equation~\ref{#1}}

\def\1{\bm{1}}

\DeclareMathAlphabet{\mathsfit}{\encodingdefault}{\sfdefault}{m}{sl}
\SetMathAlphabet{\mathsfit}{bold}{\encodingdefault}{\sfdefault}{bx}{n}

\usepackage{hyperref}
\usepackage{url}
\usepackage{enumitem}
\usepackage{graphicx}
\usepackage{booktabs}
\usepackage{tikz}
\usetikzlibrary{positioning}
\usepackage{algorithm}
\usepackage{algpseudocode}
\usepackage{tabularx}
\usepackage{xcolor}
\usepackage{array}

\title{PROTEAN: Beyond De Novo Generation --- Can {LLM} Agents Perform Text-Guided Protein Structure Editing?}

\author{%
\begin{minipage}[t]{\dimexpr\textwidth-2\tabcolsep\relax}
\centering
\normalfont\small
Encheng Su\textsuperscript{1,6,*}, Xinping Liu\textsuperscript{2,6,*},
Yizhou Wang\textsuperscript{3}, Bozitao Zhong\textsuperscript{4},\\[2pt]
Pengze Li\textsuperscript{2,6}, Jianyu Wu\textsuperscript{4,6},
Mingxiao Li\textsuperscript{6}, Xinzhu Ma\textsuperscript{5},\\[2pt]
Aoran Wang\textsuperscript{6}, Liang Hong\textsuperscript{4},
Wanli Ouyang\textsuperscript{6}, Chen Tang\textsuperscript{3,6,$\dagger$},\\[2pt]
Houqiang Li\textsuperscript{1}, Shixiang Tang\textsuperscript{3,6,$\dagger$}\\[7pt]
\footnotesize
\textsuperscript{1}University of Science and Technology of China\\
\textsuperscript{2}Fudan University\\
\textsuperscript{3}The Chinese University of Hong Kong\\
\textsuperscript{4}Shanghai Jiao Tong University\\
\textsuperscript{5}Beihang University\\
\textsuperscript{6}Shanghai Artificial Intelligence Laboratory\\[5pt]
\textsuperscript{*}Equal contribution.\quad
\textsuperscript{$\dagger$}Corresponding authors.
\end{minipage}%
}

\newcommand{\projName}[0]{\textsc{Protean}}

\iclrfinalcopy

\begin{document}

\maketitle

\fancyhead{}

\begingroup
\renewcommand{\thefootnote}{}
\footnotetext{We will continue to improve PROTEAN and expand its experimental evaluation.}
\endgroup

\begin{abstract}
Generative AI has achieved remarkable success in \emph{de novo} protein design, yet protein engineering typically begins with existing, characterized proteins.
The primary demand is targeted local editing, e.g., engineering a specific region to optimize affinity, alter immunogenicity, or improve stability, while strictly preserving core catalytic motifs and overall scaffold integrity.
However, no standardized framework currently exists for carrying out and evaluating such edits.
In current generative-design workflows, researchers often need to manually specify complex low-level parameters and rely on repetitive, trial-and-error sampling and evaluation across generative backends to search for viable conformations. While large language models (LLMs) offer an intuitive interface to express these goals using natural language, prevailing structure-generation models cannot directly interpret high-level human instructions.
To bridge this gap, we introduce \projName, a standardized benchmark for text-guided protein structural editing. \projName{} comprises 380 sub-tasks spanning six intervention families and evaluates a two-stage pipeline in which an LLM translates natural-language goals into backend-specific geometric constraints and a generative model remodels the protein backbone. A two-tier evaluation separates geometric compliance, including target attainment and off-target scope fidelity, from biophysical realizability assessed by inverse folding and refolding consistency. Across five frontier LLMs paired with three backends, strict end-to-end biophysical success starts at just 7.0\%. Ablation studies show that LLM constraint compilation is already highly accurate, whereas the main bottlenecks are backend-specific scope leakage and refolding failure. \projName{} provides an open testbed for developing autonomous, multi-backend macromolecular design agents.
\end{abstract}

\section{Introduction}
\label{sec:intro}


Artificial intelligence has fundamentally transformed structural biology. Deep learning systems such as AlphaFold2 and AlphaFold3 have turned proteome-wide structure prediction into standard practice~\citep{jumper2021highly,abramson2024accurate}, while 3D generative architectures like RFdiffusion~\citep{watson2023de}, Chroma~\citep{ingraham2023illuminating}, and ESM3~\citep{hayes2025simulating} have unlocked programmable \emph{de novo} backbone generation from scratch. These generative frameworks, however, operate primarily in an unconstrained regime—synthesizing full proteins from stochastic noise or unconstrained boundary priors.

In real-world protein engineering, however, projects rarely start from scratch. Practitioners almost invariably build upon existing, evolutionarily vetted proteins to inherit their natural stability, solubility, and proven expression profiles—qualities that \emph{de novo} designs frequently lack. 
Across both therapeutic development and industrial biocatalysis, the primary demand is \emph{targeted structural editing}: remodeling a specific local region to optimize affinity, alter immunogenicity, or improve stability, while strictly preserving core catalytic motifs and overall scaffold integrity. 
Yet, executing targeted edits computationally remains an empirical ordeal. 
Because existing protein models provide limited localized control, often permitting unintended structural drift outside the target site, practitioners must manually configure low-level parameters and rely on repetitive, trial-and-error sampling across fragmented backends. 

Large language models (LLMs) provide an intuitive interface to streamline this process, allowing engineers to express nuanced structural objectives in natural language. 
Yet, prevailing \textbf{3D generative engines cannot directly parse high-level human instructions}; they require specialized, low-level geometric constraints and heterogeneous parameter configurations. 
Bridging this gap generally demands an two-stage LLM-plus-engine collaborative paradigm: an upstream reasoning LLM translates abstract natural-language instructions into backend-specific geometric constraints, and a downstream generative backend executes the physical 3D backbone remodeling. While prior biology-oriented language models have addressed sequence-level editing or tool invocation for \emph{de novo} design~\citep{liu2024a,weng2025disprotedit,ghafarollahi2024protagents}, no standardized benchmark currently exists to formalize, execute, and evaluate text-guided 3D backbone editing under strict conservation contracts across diverse generative backends.

To establish a principled foundation for this emerging problem, we introduce \projName, the first standardized benchmark for text-guided protein structural editing. \projName establishes 380 tasks spanning eight structural intervention families: secondary-structure shifting (helix and sheet enrichment), inter-domain distance modulation (approach and separation), reference-fragment mimicry, global shape control, backbone cyclization, and motif scaffolding. 
These tasks collectively evaluate an agent's ability to drive purposeful deformation while preserving non-target regions under a rigid geometric contract 
($C_\alpha\text{-RMSD} \le 0.5\,\text{\AA}$). 

Crucially, \projName institutes a rigorous two-tier evaluation protocol that decouples geometric compliance from biophysical realizability:
\textbf{Tier 1: Geometric Compliance} verifies macroscopic task execution directly on the generated backbone, measuring both target attainment (whether the instructed structural transformation succeeded) and off-target scope fidelity (whether unedited scaffolds strictly satisfied the conservation threshold); 
\textbf{Tier 2: Biophysical Realizability} subjects geometrically compliant backends to sequence redesign via ProteinMPNN~\citep{dauparas2022robust} and orthogonal refolding via ESMFold~\citep{lin2023evolutionary}, examining whether the computationally edited backbone can stably self-assemble in physical solution.

\begin{figure}[t]
    \centering
    \includegraphics[width=1\linewidth]{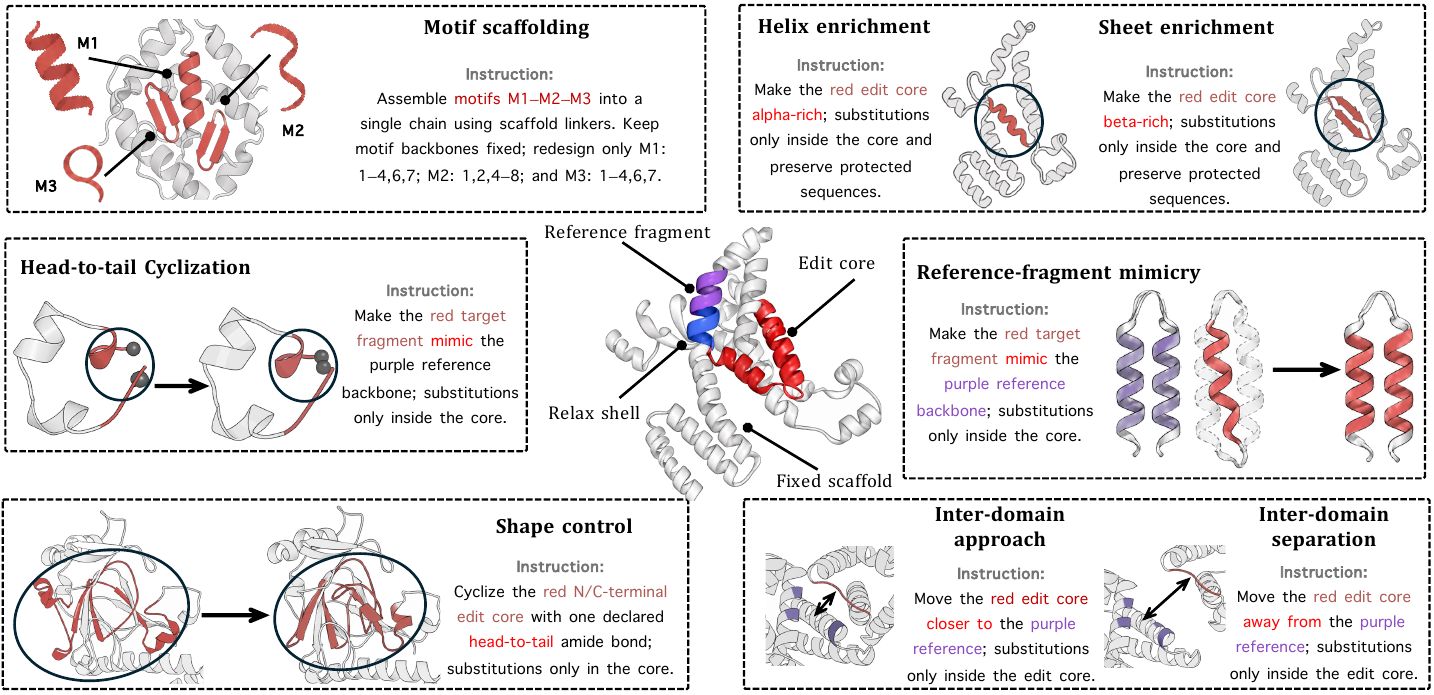}
    \vspace{-0.5cm}
    \caption{\textbf{Overview of the eight protein structure editing tasks in \textsc{Protean}.}
    The benchmark covers motif scaffolding, helix enrichment, sheet enrichment, head-to-tail cyclization, reference-fragment mimicry, shape control, inter-domain approach, and inter-domain separation. Across the tasks, red denotes the editable core, blue denotes the surrounding relaxation shell, gray denotes the fixed scaffold, and purple denotes the reference fragment or region. Representative structures and natural-language instructions illustrate the editing objectives and the associated sequence and geometric constraints.}
    \vspace{-0.5cm}
    \label{fig:protean_tasks}

\end{figure}

\projName{} fundamentally redefines computational protein editing along two core dimensions. 
First, it transforms targeted structural editing from an \emph{ad hoc}, trial-and-error craft into a formalized task governed by rigid geometric conservation contracts. 
Second, it establishes natural language as an expressive, high-level control interface, thereby democratizing macromolecular design by empowering researchers to direct complex, multi-backend 3D remodeling through scientific intent rather than fragile parameter hacking. (Fig.~\ref{fig:protean_tasks})
Beyond its utility as an execution framework, \projName{} provides the foundational diagnostic testbed required to rigorously interrogate and catalyze next-generation, physically realizable autonomous design agents.

Through comprehensive benchmarking across five frontier LLMs paired with three heterogeneous generative backends (BoltzGen~\citep{stark2025boltzgen}, ODesign~\citep{zhang2025odesign}, and RFdiffusion3~\citep{butcher2025rfd3}), our experiments uncover that autonomous structural editing is viable yet severely bottlenecked. 
Specifically, even the top-performing agent (Claude-Opus-5) attains only 46.0\% joint geometric compliance ($G_1 \wedge G_3$, Table~\ref{tab:family-main}) with competitor baselines falling below 28\%, while achieving end-to-end biophysical refolding self-consistency poses an even steeper sampling barrier (Table~\ref{tab:motif-bounds}).

Systematic bottleneck analyses reveal that this performance collapse stems not from semantic misunderstanding, but from generator-specific failures, including pervasive non-local contact strain, severe scope leakage (collateral drift in designated invariant scaffolds), and persistent Pareto deadlocks during complex remodeling. 
This work delivers three primary contributions:
\begin{enumerate}
\item We formalize text-guided protein structural editing under hard conservation boundaries, establishing a two-stage collaborative paradigm that connects high-level natural language to low-level 3D generative backends across eight structural intervention families.
\item We introduce \projName, comprising 380 verified editing sub-tasks across six intervention families and eight concrete tasks. It is paired with an evaluation harness that decouples macroscopic geometric compliance from microscopic biophysical realizability.
\item We provide extensive empirical evaluations of a set of paired agent-backend toolchains, isolating backend coordinate leakage and refolding collapse as the primary bottlenecks, thereby offering clear directions for developing autonomous macromolecular design agents.
\end{enumerate}

\section{Related Work}
\label{sec:related}

\subsection{Structure Generation and Heterogeneous Design Backends}
Recent structural generative models~\citep{watson2023de,ingraham2023illuminating,hayes2025simulating} have demonstrated remarkable capabilities in \emph{de novo} backbone synthesis and motif scaffolding. However, these systems rely heavily on human experts manually constructing complex, backend-specific inputs. Moreover, current state-of-the-art generators exhibit fragmented, specialized operational domains rather than interchangeable capabilities: BoltzGen~\citep{stark2025boltzgen} natively handles topological head-to-tail cyclization through explicit covalent bond constraints; ODesign~\citep{zhang2025odesign} excels at partial diffusion and hotspot-conditioned inter-domain distance modulation; and RFdiffusion3 provides robust generalized backbone generation under spatial coordinate bounds. 
\textsc{Protean} treats these heterogeneous engines as frozen tools, evaluating how effectively LLMs can autonomously select and compile their incompatible native specifications. 

\subsection{Language-Driven Protein Engineering}
A growing body of work leverages natural language to steer macromolecular generation. 
Although existing language-driven methods~\citep{liu2024a, weng2025disprotedit, wang2024instructprotein, lv2024prollama} edit 1D protein sequences, text-guided 3D structure editing remains the core imperative for functional design. Operating solely on sequences neglects explicit geometric constraints, offers no spatial preservation for unedited regions, and fails to yield native 3D representations. 
On the other hand, scientific agent frameworks such as ScienceAgentBench~\citep{chen2025scienceagentbench}, ChemCrow~\citep{bran2024augmenting}, and BioDesignBench~\citep{kim2026biodesignbench} evaluate tool-use workflows, but predominantly target general coding or \emph{de novo} binder generation pipelines. 


\subsection{Protein Design Benchmarks \& Decoupled Evaluation}
Existing protein design benchmarks focus on isolated stages, such as representation learning (e.g., TAPE, PEER)~\citep{rao2019evaluating,xu2022peer}, fitness prediction (ProteinGym)~\citep{notin2023proteingym}, or inverse folding (ProteinInvBench)~\citep{gao2023proteininvbench}. Consequently, they fail to diagnose the multi-tier failure modes of text-guided editing, often conflating geometric non-compliance with biophysical instability under a monolithic score. \textsc{Protean} bridges this gap by introducing a decoupled evaluation protocol that rigorously disentangles 3D structural geometric fidelity from downstream biophysical plausibility, establishing a fine-grained diagnostic benchmark for autonomous protein engineering.

\section{The \projName{} Benchmark}
\label{sec:benchmark}




\projName{} evaluates a model's ability to perform targeted structural interventions on an existing protein while strictly preserving its unedited scaffold, with natural language. 
The benchmark comprises eight task categories (380 sub-tasks): seven length-preserving structure-editing tasks ($350$ sub-tasks) and one motif-scaffolding task ($30$ sub-tasks) governed by a variable-length output contract. 

\subsection{Task Formulation}
\label{sec:task}

A \projName{} task is formally defined as a tuple:
\begin{equation}
\tau =
\left(
    \underbrace{(s,X)}_{\text{source protein}},
    \underbrace{\mathcal{M}}_{\text{edit scope}},
    \underbrace{u}_{\text{instruction}}
\right),
\qquad
s \in \mathcal{A}^{L},
\quad
X \in \mathbb{R}^{L \times 4 \times 3}, \nonumber
\end{equation}
where $s$ denotes the amino-acid sequence of length $L$ over the canonical alphabet $\mathcal{A}$, $X$ contains the backbone heavy-atom coordinates $(\mathrm{N}, \mathrm{C}_\alpha, \mathrm{C}, \mathrm{O})$, $u$ is a natural-language editing instruction, and $\mathcal{M}=(\mathcal{M}_{\mathrm{edit}}, \mathcal{M}_{\mathrm{relax}}, \mathcal{M}_{\mathrm{prot}})$ partitions the residues into editable, flexible (relaxation-allowed), and strictly protected regions.

For the seven length-preserving tasks, valid solutions must retain the source sequence length, chain topology, and canonical residue ordering (i.e., insertions, deletions, and non-canonical amino acids are strictly disallowed). Sequence substitutions are restricted exclusively to $\mathcal{M}_{\mathrm{edit}}$, whereas residues in $\mathcal{M}_{\mathrm{prot}}$ must preserve their original identities. Spatially, after anchored superposition onto the scaffold, all backbone atoms in $\mathcal{M}_{\mathrm{prot}}$ must exhibit an RMSD within $0.1$~\AA{} relative to the source.

By enforcing an explicit edit scope, sequence constraints, and deterministic success criteria, \projName{} strictly demarcates targeted \emph{structural editing} from unconstrained regeneration: achieving a design objective by reconstructing unperturbed regions violates the preservation contract. \projName{} thus benchmarks whether an agent or model can reliably ground language-guided structural interventions into executable design pipelines under stringent structural preservation constraints. This complements existing benchmarks in structure prediction, inverse folding, and fitness prediction \citep{Rao2019,Notin2023}, text-guided sequence design \citep{Liu2025}, and constrained generative models such as RFdiffusion, BoltzGen, and ODesign.

\subsection{Task Taxonomy and Criteria}
\label{sec:edit-types}


To systematically probe localized structural control, we formulate six intervention families across eight concrete tasks that span orthogonal biophysical dimensions, from local secondary structure and core packing to inter-domain kinematics, chain topology, and functional motif scaffolding. 
Evaluation is grounded in reproducible structural contracts: task success verifies that a candidate strictly satisfies the prescribed geometric constraints.

\paragraph{Secondary-structure enrichment (2 tasks).}
These tasks require remodeling the editable region to become either $\alpha$-helix-rich (\textbf{50 sub-tasks}) or $\beta$-strand-rich (\textbf{50 sub-tasks}). Secondary structures are assigned via DSSP \citep{KabschSander1983,Touw2015}, and the objective is quantified by the propensity contrast $\Delta f = f_{\mathrm{target}} - f_{\mathrm{opposing}}$. Passing requires $\Delta f \ge 0.35$ for helix enrichment and $\Delta f \ge 0.10$ for sheet enrichment. These absolute thresholds evaluate whether a model can actively reshape local hydrogen-bonding networks and steric rigidity, rather than merely generating sequences statistically biased toward a particular fold.

\paragraph{Inter-domain distance modulation (2 tasks).}
These tasks evaluate spatial repositioning by moving an editable structural element either toward (\emph{approach}, \textbf{50 sub-tasks}) or away from (\emph{separation}, \textbf{50 sub-tasks}) a protected reference patch. In a shared global frame aligned to the source scaffold, success requires a median distance change of at least $1.0$~\AA{} (decreasing for approach, increasing for separation; see Table~\ref{tab:edit-types}). This abstracts conformational switching—such as allosteric domain closure or relief of autoinhibition \citep{ChangeuxEdelstein2005}—into a localized hinge- or loop-remodeling problem under strict contextual preservation.

\paragraph{Reference-fragment mimicry (1 task).}
Given a target backbone fragment and explicit residue mappings, the model must remodel the editable region to reproduce the prescribed geometry (\textbf{50 sub-tasks}). A design passes if the mapped backbone RMSD is $\le 2.0$~\AA{}, covering $\ge 80\%$ of the designated alignment with at least three aligned residues. This task benchmarks structural transplantation, formalizing practical workflows such as CDR loop grafting \citep{Jones1986}, catalytic-site transfer, and epitope presentation \citep{Ofek2010} within a fixed scaffold.

\paragraph{Core burial (1 task).}
Starting from an exposed 12-residue window, the model must pack the segment into a more buried and compact microenvironment (\textbf{50 sub-tasks}). Passing criteria require the median relative solvent accessibility (RSA, normalized via SASA \citep{Tien2013}) to decrease by at least $0.10$ to an absolute $\mathrm{RSA} \le 0.40$, alongside bounded variations in the local radius of gyration. This reflects hydrophobic core repacking—the primary thermodynamic driver of protein folding and stability \citep{Chothia1984,Rose1985}.

\paragraph{Head-to-tail cyclization (1 task).}
Cyclization tasks define the protein termini as editable and require linking the N- and C-termini through a peptide bond (\textbf{50 sub-tasks}). A valid edit must form exactly one new covalent bond with a terminal backbone $\mathrm{C}\text{--}\mathrm{N}$ distance between $1.20$ and $1.50$~\AA{}. Unlike purely geometric alterations, cyclization modifies chain topology; it evaluates whether a model can perform topological closure to enhance proteolytic resistance or preorganize binding conformations \citep{Rettie2025a,Rettie2025b} while maintaining scaffold integrity.

\paragraph{Motif scaffolding (1 task).}
Operating under a variable-length contract, motif scaffolding constructs a new single-chain scaffold around one or more disconnected functional segments (e.g., catalytic motifs, metal-binding sites), this task has 30 sub-tasks. Following MotifBench \citep{Zheng2025}, success requires motif backbone preservation ($\mathrm{RMSD} \le 1.0$~\AA{}) and computational self-consistency: at least one inverse-folded sequence must refold into the designed structure. This serves as the structural dual of targeted editing: holding the functional geometry invariant while rebuilding the entire structural carrier \citep{Watson2023}.

\subsection{Benchmark Curation Pipeline}
\label{sec:construction}

\begin{figure}
    \centering
    \includegraphics[width=1\linewidth]{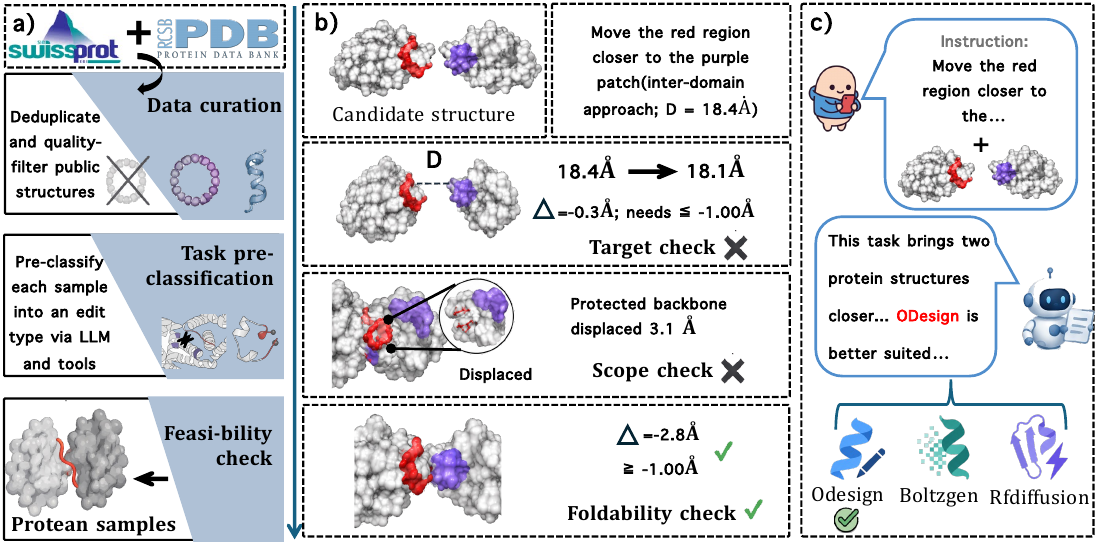}
    \vspace{-0.5cm}
    \caption{\textbf{Overview of the \textsc{Protean} benchmark construction and model-routing pipeline.}
    \textbf{(a)} Protein structures are collected from Swiss-Prot and the PDB, deduplicated, quality-filtered, pre-classified by edit type, and screened for feasibility.
    \textbf{(b)} Candidate edits are validated through target-achievement, scope compliance, and refolding self-consistency checks to retain feasible samples.
    \textbf{(c)} Given a protein structure and a natural-language instruction, the model interprets the editing task and selects an appropriate backend, such as ODesign, BoltzGen, or RFdiffusion.}
    \vspace{-0.5cm}
\label{fig:protean_pipeline}

    \label{fig:figure2}
\end{figure}

As illustrated in Figure~\ref{fig:figure2}, \projName{} instances are constructed via a three-stage curation funnel:

\textbf{Data curation.} Starting from public protein structures PDB \citep{Berman2000}, we build a high-quality, non-redundant source pool. Sequence deduplication and family-based clustering prevent homologous structures from entering the benchmark as spurious independent instances (detailed filtering thresholds are provided in Appendix~\ref{app:pipelinesource} and Appendix~\ref{app:motif}). 
The motif-scaffolding task is adopted directly from MotifBench \citep{Zheng2025}. 

\textbf{Task pre-classification.} We then evaluate each curated structure's suitability across edit families. By coupling LLM reasoning with structural analysis tools, this stage assigns candidate edit types and identifies the requisite structural context (e.g., editable vs. protected regions, reference inputs). This tailors general proteins into well-defined, type-specific editing problems rather than treating them interchangeably across tasks.

\textbf{Feasibility check.} Finally, candidate problems are filtered against the predefined constraints of their assigned edit type. Only validated source--task pairs are retained. The resulting \projName{} benchmark comprises $380$ equal-length editing tasks ($50$ for each of the seven edit types and $30$ motif-scaffolding cases). 
To minimize potential protein-data leakage—particularly the possibility that an LLM may have encountered related protein information during pretraining—we applied a protein-blinding procedure (see Appendix~ \ref{app:pipelineblind}). 

\subsection{Multi-Gated Evaluation Protocol}
\label{sec:evaluation}



We evaluate protein editing across three complementary dimensions, formalized as the corresponding validation gates:

\paragraph{Target attainment ($G_1$).}
Assesses whether the intended structural modification is achieved according to type-specific criteria (Table~\ref{tab:edit-types}). All evaluation thresholds are frozen prior to model inference. For example, an inter-domain approach task is claimed to pass the gate only when its source-referenced median distance decreases by at least $1.0$~\AA{}

\paragraph{Refolding self-consistency ($G_2$).}
Verifies the biophysical plausibility of the edited backbone via a fixed redesign-and-refolding loop. 
For length-preserving tasks, the editable region is redesigned with ProteinMPNN~\citep{Dauparas2022} while preserving the protected context. Four candidate sequences are folded using ESMFold~\citep{Lin2023}; passing requires at least one sequence to meet frozen whole-chain and protected-context RMSD thresholds. This serves as a finite-budget proxy, meaning failure denotes a sampling miss rather than definitive biological incompatibility. (Cyclization is evaluated via a separate protocol.)

\paragraph{Scope compliance ($G_3$).}
Ensures the output is structurally sound and strictly constrained within the declared edit boundaries. Protected sequence identities, residue mapping, and backbone geometry must adhere to benchmark specifications. Candidates that fulfill target criteria by altering off-target regions are deemed invalid.

Overall success requires passing all applicable gates. Furthermore, execution outcomes (e.g., malformed outputs, backend failures) are reported separately to avoid conflating runtime anomalies with algorithmic editing failures (see Appendix~\ref{app:gatepseudo} for full scoring details).

\section{Experiments}
\label{sec:experiments}

\subsection{Experimental Setup}
We evaluate five frontier large language models: Claude-Opus-5~\citep{anthropic2025claude}, GPT-5.6-Sol~\citep{openai2025gpt}, DeepSeek-V4-Pro~\citep{deepseek2025deepseek}, Qwen3.8Max~\citep{qwen2025qwen}, and Kimi-K3~\citep{moonshot2025kimi}, they act as autonomous constraint compilers across three structurally heterogeneous generation engines: BoltzGen, ODesign, and RFdiffusion3. Evaluations are conducted on the standardized \textsc{Protean} suite curated in Section~\ref{sec:benchmark}, comprising $380$ tasks across the two complementary paradigms: $350$ length-preserving editing tasks ($50$ instances across each of the seven structural intervention types) and $30$ de novo motif-scaffolding cases, evaluated strictly under the formulation and criteria established in Sections~\ref{sec:task} and~\ref{sec:edit-types}.

\paragraph{Evaluation Protocol.}

\noindent We use pass@K as the evaluation metric. Furthermore, we tailor evaluation criteria to the distinct physical nature of each task:
\begin{itemize}[leftmargin=*, nosep]
    \item \textbf{Structural Editing ($\mathcal{G}_{\text{edit}} = G_1 \wedge G_3$):} 
    Because equal-length editing operates on native protein contexts, the central challenge is executing local deformations without corrupting global structural integrity. We thus assess task success via \emph{Joint Geometric Compliance}, requiring designs to concurrently fulfill target modifications (\emph{Target Attainment}, $G_1$) and respect off-target boundary constraints (\emph{Scope Compliance}, $G_3$). Refolding self-consistency ($G_2$) is decoupled and analyzed as a downstream sampling bottleneck in Appendix~\ref{app:bottlenecks}.

    \item \textbf{Motif Scaffolding ($\mathcal{G}_{\text{motif}} = G_1 \wedge G_2 \wedge G_3$):} 
    In contrast, motif scaffolding generates backbones \emph{de novo}. Lacking native scaffolds to ensure refolding self-consistency, purely geometric criteria risk admitting physically unrealizable artifacts. Adhering to community benchmarks~\citep{zheng2025motifbench}, we mandate orthogonal biophysical verification for candidate validity: backbones must preserve motif geometry ($G_1 \wedge G_3$) and produce at least one sequence via ProteinMPNN that refolds via ESMFold with whole-chain $\text{C}_\alpha\text{-RMSD} < 2.0\,\text{\AA}$ and motif $\text{C}_\alpha\text{-RMSD} < 1.0\,\text{\AA}$ ($G_2$).
\end{itemize}

\definecolor{top1}{HTML}{EAD3D5} 
\definecolor{top2}{HTML}{D4E1E8} 
\newcommand{\toptwo}[1]{\colorbox{top2}{\hspace{-1.5pt}#1\hspace{-1.5pt}}}
\newcommand{\topone}[1]{\colorbox{top1}{\hspace{-1.5pt}#1\hspace{-1.5pt}}}

\begin{table*}[t]
\centering
\caption{Per-task gate-wise pass rates across 7 in-scaffold editing tasks (pass@16, best backend). $G_1$: target attainment, $G_3$: scope compliance, $\wedge$: joint $G_1\wedge G_3$. Thresholds: Approach/Separation ($\Delta \mathrm{dist} \ge 1.0$~\AA), Mimicry ($\mathrm{RMSD} \le 2.0$~\AA, cov $\ge 80\%$), Core burial ($\Delta\mathrm{RSA} \ge 0.10, \mathrm{RSA} \le 0.40$), Cyclization ($\mathrm{C\text{--}N} \in [1.2, 1.5]$~\AA). \colorbox{top1}{Red} and \colorbox{top2}{blue} denote highest and second-highest scores per task.}
\label{tab:family-main}
\footnotesize 
\setlength{\tabcolsep}{1.6pt} 
\setlength{\fboxsep}{1.0pt}   
\renewcommand{\arraystretch}{1.15}
\begin{tabular}{@{}l ccc ccc ccc ccc ccc@{}}
\toprule
& \multicolumn{3}{c}{\textbf{DeepSeek-V4-Pro}} & \multicolumn{3}{c}{\textbf{Qwen3.8Max}} &
  \multicolumn{3}{c}{\textbf{Claude-Opus-5}} & \multicolumn{3}{c}{\textbf{GPT-5.6-Sol}} &
  \multicolumn{3}{c}{\textbf{Kimi-K3}} \\
\cmidrule(lr){2-4}\cmidrule(lr){5-7}\cmidrule(lr){8-10}\cmidrule(lr){11-13}\cmidrule(lr){14-16}
\textbf{Task / Family} & $G_1$ & $G_3$ & $\wedge$ & $G_1$ & $G_3$ & $\wedge$ & $G_1$ & $G_3$ & $\wedge$ & $G_1$ & $G_3$ & $\wedge$ & $G_1$ & $G_3$ & $\wedge$ \\
\midrule
\multicolumn{16}{@{}l}{\textit{Secondary-structure enrichment}} \\
\quad $\alpha$-Helix & 86 & \topone{100} & 66.0 & 88 & \topone{100} & 58.0 & \topone{98} & \topone{100} & \topone{90.0} & \toptwo{96} & \topone{100} & \toptwo{84.0} & 88 & \topone{100} & 76.0 \\
\quad $\beta$-Sheet  & 62 & \topone{94}  & 44.0 & 60 & \topone{94}  & 36.0 & \topone{84} & \topone{94}  & \topone{60.0} & \toptwo{82} & \topone{94}  & \toptwo{58.0} & 60 & \topone{94}  & 40.0 \\
\addlinespace[2pt]
\multicolumn{16}{@{}l}{\textit{Inter-domain distance modulation}} \\
\quad Approach   & 36 & \topone{100} & 18.0 & 38 & \topone{100} & 12.0 & \topone{56} & \topone{100} & \topone{36.0} & 46 & \topone{100} & \toptwo{26.0} & \toptwo{48} & \topone{100} & 14.0 \\
\quad Separation & 46 & \topone{100} & 20.0 & 44 & \topone{100} & 12.0 & 48 & \topone{100} & \topone{34.0} & \topone{52} & \topone{100} & \toptwo{28.0} & \toptwo{50} & \topone{100} & 18.0 \\
\addlinespace[2pt]
\multicolumn{16}{@{}l}{\textit{Localized structural transplantation \& repacking}} \\
\quad Fragment mimicry & 18 & \topone{78}  & 16.0 & 18 & \topone{78}  & 22.0 & \topone{76} & \topone{78}  & \topone{56.0} & \toptwo{58} & \topone{78}  & \toptwo{38.0} & 18 & \topone{78}  & 14.0 \\
\quad Core burial      & 34 & \topone{96}  & 30.0 & 36 & \topone{96}  & 28.0 & \topone{66} & \topone{96}  & \topone{46.0} & \toptwo{50} & \topone{96}  & \toptwo{34.0} & 34 & \topone{96}  & 20.0 \\
\addlinespace[2pt]
\multicolumn{16}{@{}l}{\textit{Chain topological alteration}} \\
\quad Cyclization      & \topone{100} & \topone{100} & 0.0 & \topone{100} & \topone{100} & 0.0 & \topone{100} & \topone{100} & 0.0 & \topone{100} & \topone{100} & 0.0 & \topone{100} & \topone{100} & 0.0 \\
\midrule
\textbf{Overall}       & 54.6 & \topone{95.4} & 27.7 & 54.9 & \topone{95.4} & 24.0 & \topone{75.4} & \topone{95.4} & \topone{46.0} & \toptwo{69.1} & \topone{95.4} & \toptwo{38.3} & 56.9 & \topone{95.4} & 26.0 \\
\bottomrule
\end{tabular}
\vspace{-0.5cm}
\end{table*}

\subsection{Structural Geometric Compliance Across Task Families}
\label{sec:structural-compliance}

\paragraph{Macro Stratification and the Joint-Gate Imperative.}
As detailed in Table~\ref{tab:family-main}, evaluating structural editing across frontier reasoning agents reveals distinct capability tiers under the joint compliance criterion ($G_1 \wedge G_3$ at pass@16). 
Claude-Opus-5 defines the benchmark frontier (46.0\% overall), followed by GPT-5.6-Sol (38.3\%), while remaining models cluster substantially lower (24.0--27.7\%). 
Crucially, isolated target attainment ($G_1$) or boundary compliance ($G_3$) paints a deceptively optimistic picture: marginal $G_3$ rates saturate across all architectures due to pass@16 sampling diversity. 
The actual spatial reasoning bottleneck appears exclusively under the joint gate ($G_1 \wedge G_3$), which penalizes agents that satisfy design goals at the expense of unedited scaffold integrity within the \emph{identical} candidate (Appendix~\ref{app:case-studies}).

\paragraph{Contact Order Governs Secondary-Structure Remodeling.}
Performance on secondary-structure enrichment is fundamentally bounded by physical contact order. 
Helix design consistently outperforms sheet design across all models by 20--30 percentage points. 
Because $\alpha$-helices depend on strictly local, contiguous hydrogen bonds ($i \to i+4$), generative backends nucleate them readily without disrupting tertiary topology. 
Conversely, $\beta$-sheet enrichment requires coordinating non-local strand pairing and registry with the preserved framework. 
Generative backends repeatedly deform flanking conserved residues to satisfy inter-strand geometries, triggering pervasive scope violations.

\paragraph{Reference Mimicry as the Benchmark Discriminator.}
Fragment mimicry exhibits the steepest performance dispersion, spanning a four-fold divergence between Claude-Opus-5 (56.0\%) and Kimi-K3 (14.0\%). 
Mimicry represents the benchmark's most demanding test of spatial reasoning, requiring models to simultaneously solve discrete residue-index mapping and continuous rigid-body superposition ($\mathrm{RMSD} \le 2.0$~\AA{} over $\ge 80\%$ coverage) under invariant boundary bounds. 
Sub-frontier models predominantly fail at instruction translation, either emitting misaligned residue mappings or proposing structurally incompatible grafts whose insertion strain shatters contextual tolerances.

\paragraph{Mechanical Strain in Relational and Compaction Edits.}
Inter-domain modulation (Approach/Separation) and Shape control (Core Burial) both expose the mechanical tension between metric optimization and coordinate invariance. 
In relational modulation, baseline models frequently command rigid-body translations without designing compensating flexibility into transition hinges, transmitting excessive torque into the protected core. 
Similarly, driving core burial via hydrophobic compaction ($\Delta\mathrm{RSA} \ge 0.10$ to $\le 0.40$) inherently exerts compressive forces on the backbone, creating structural friction that severely caps joint compliance across all architectures.

\paragraph{The Pareto Deadlock of Topological Cyclization.}
Head-to-tail cyclization uncovers an absolute Pareto incompatibility: while all models reach 100\% pass rates on marginal $G_1$ (forming the peptide bond) and $G_3$ (boundary preservation), joint compliance collapses completely to \textbf{0.0\%} ($G_1 \wedge G_3$). 
Under rigid Euclidean bounds, these objectives are mutually exclusive: closing the separated termini ($1.20\text{--}1.50$~\AA{}) inevitably causes global backbone relaxation that breaches $G_3$, whereas strictly pinning scaffold coordinates immobilizes the termini ($G_1$ failure). 
This universal 0.0\% impasse demonstrates that cyclic topological edits demand concerted multi-scale deformation planning rather than isolated localized re-sampling (Appendix~\ref{app:cyclization-decomposition}).

\begin{figure}
    \centering
    \includegraphics[width=1\linewidth]{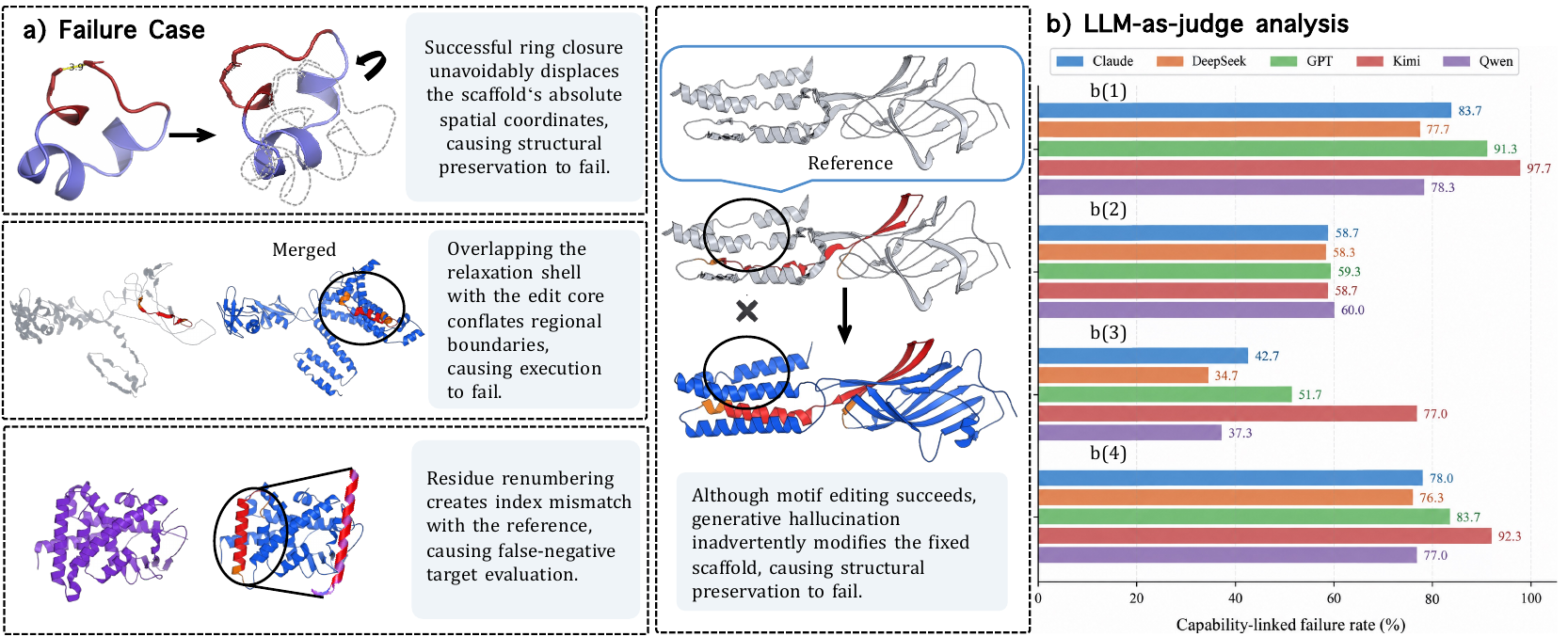}
    \vspace{-0.5cm}
    \caption{\textbf{Failure mode analysis and LLM-as-judge capability evaluation in ODesign.} 
\textbf{(a) Representative failure cases in protein editing.} 
Detailed mechanistic breakdowns of structural and operational failures, including absolute coordinate displacement during ring closure, scaffold modifications induced by generative hallucination, regional boundary conflation from overlapping relaxation shells, and evaluation mismatches caused by residue renumbering. 
\textbf{(b) Capability-linked failure profiling via LLM-as-judge.} 
Quantitative evaluation across leading LLMs categorized into four distinct failure modes: 
\textit{(1) Edit-intent translation failure}, 
\textit{(2) Spatial-relation planning failure}, 
\textit{(3) Structural-boundary preservation failure}, and 
\textit{(4) Multi-constraint coordination failure}. 
Unsupported tasks and out-of-memory (OOM) instances are excluded; percentages indicate capability-linked failure rates and may overlap across categories.}
\vspace{-0.5cm}
\label{fig:failure_analysis}
\end{figure}


\begin{table}[t]
\centering
\caption{Motif scaffolding performance comparing frontier model ceilings against competitor baselines across pipelines (ODesign, Boltzgen, RFdiffusion3). Values denote success rates (\%) under strict biophysical refolding self-consistency ($G_1 \wedge G_2 \wedge G_3$). \colorbox{top1}{Red} and \colorbox{top2}{blue} highlights indicate the highest and second-highest scores per column, respectively.}
\label{tab:motif-bounds}
\small
\setlength{\tabcolsep}{9pt}
\setlength{\fboxsep}{1.8pt}
\renewcommand{\arraystretch}{1.25}
\begin{tabular}{@{}l ccccc@{}}
\toprule
\textbf{Model} & \textbf{$s@1$} & \textbf{$s@2$} & \textbf{$s@4$} & \textbf{$s@8$} & \textbf{$s@16$} \\
\midrule
Claude-Opus-5    & \toptwo{13.0} & \toptwo{20.0} & \toptwo{23.0} & \topone{40.0} & \topone{47.0} \\
GPT-5.6-Sol      & \topone{17.0} & \topone{23.0} & \topone{27.0} & \topone{40.0} & \toptwo{43.0} \\
DeepSeek-V4-Pro  & \toptwo{13.0} & 17.0          & 20.0          & 27.0          & 33.0          \\
Qwen3.8Max       & 10.0          & \toptwo{20.0} & 20.0          & 23.0          & 23.0          \\
Kimi-K3          &  7.0          & 13.0          & 20.0          & 23.0          & 27.0          \\
\bottomrule
\end{tabular}
\vspace{-0.5cm}
\end{table}

\subsection{Functional Motif Scaffolding and Biophysical Realizability}
\label{sec:motif-analysis}

\paragraph{The Refolding Imperative in De Novo Carrier Design.}
While fixed-scaffold editing leverages evolutionary backbones that anchor global stability, motif scaffolding requires generating entirely new single-chain carriers around discontinuous functional segments. 
In this variable-length regime, generative engines easily hallucinate backbones that satisfy static spatial alignments with the motif but represent thermodynamically unviable conformations. 
Following structural biology standards~\citep{Zheng2025}, Table~\ref{tab:motif-bounds} mandates orthogonal computational refolding verification ($G_2$, via sequence redesign with ProteinMPNN and forward folding with ESMFold). 
A design is considered successful only under joint validity ($G_1 \wedge G_2 \wedge G_3$), filtering out non-physical hallucinations.

\paragraph{Frontier Ceilings and Linker Arithmetic.}
Under the full candidate budget ($s@16$), Claude-Opus-5 and GPT-5.6-Sol establish the generative frontier (43.0--47.0\% self-consistent success), substantially outpacing competitor baselines (23.0--33.0\%). 
In motif scaffolding, the reasoning agent must deduce the length partitioning of inter-motif linkers while satisfying global sequence constraints. 
Frontier models excel at inferring linker lengths that naturally accommodate canonical secondary-structure elements, thereby relieving torsional strain. 
In contrast, baseline models frequently prescribe over-compacted or excessively extended linkers that induce severe conformational tension, ultimately precipitating refolding collapse.

\paragraph{Further implementation details in motif scaffolding}
Because of the different nature of this task, we apply some distinct way to implement. We shall show the implementation details and further discussion in Appendix \ref{app:motif2}

\subsection{Mechanistic Failure Mode Profiling via LLM-as-Judge}
\label{sec:failure-modes}
To diagnose the bottlenecks behind failed designs, we dissect candidate breakdowns through mechanistic case analyses and LLM-as-judge capability profiling across four distinct failure modes (Fig.~\ref{fig:failure_analysis}).
Qualitative inspection confirms that errors stem from physical incompatibilities, such as ring-closure coordinate displacement, regional boundary conflation, and generative hallucinations mutating conserved scaffolds (Fig.~\ref{fig:failure_analysis}a).
Systematic evaluation reveals that while spatial planning fails uniformly across models ($\approx 60\%$), multi-constraint coordination failure ($76.3\%\text{--}92.3\%$) and intent translation failure remain the primary discriminators separating frontier models from baselines (Fig.~\ref{fig:failure_analysis}b).

\section{Conclusion}
\label{sec:conclusion}

\projName establishes a rigorous foundation for measuring autonomous macromolecular editing under strict physical constraints. By formalizing explicit boundary preservation contracts, source-isolated task splits, and multi-candidate evaluation, our work provides an operational benchmark to guide the development of next-generation biomolecular AI agents.

\section*{Reproducibility Statement}
All task definitions, canonical source coordinates, and frozen evaluation thresholds are standardized within the benchmark framework. Candidate alignments follow deterministic Kabsch superposition, sequence redesign utilizes fixed seeds via ProteinMPNN, and structural refolding is verified using ESMFold. Standard pass@$k$ metrics across candidate budgets ($k \in \{1, 2, 4, 8, 16\}$) evaluate empirical rollout success rates under uniform conditions. Detailed backend adapter schemas, compilation prompts, and evaluation procedures are documented in Appendix~\ref{app:pipeline}--\ref{app:gatepseudo}.

\section*{Ethics Statement}

This work uses publicly available datasets, benchmark materials, or interaction records in accordance with their respective licenses and terms of use. We do not intentionally collect personally identifiable information or access private intermediate system states beyond what is necessary for evaluation. Any released data, logs, or trace files should be limited to the information required for reproducibility and analysis, and residual identifiers should be removed where applicable.

\section*{AI Use Statement}
The language models evaluated in this study were accessed via standard API interfaces under fixed system prompts and deterministic decoding settings. Generative AI tools were also used for textual refinement and formatting assistance during manuscript preparation. All experimental protocols, structural scoring pipelines, and analytical evaluations were implemented and verified by the authors.

\bibliography{iclr2027_conference}

@article{jumper2021highly,
  author  = {Jumper, John and Evans, Richard and Pritzel, Alexander and Green, Tim and Figurnov, Michael and Ronneberger, Olaf and Tunyasuvunakool, Kathryn and Bates, Russ and {\v{Z}}{\'\i}dek, Augustin and Potapenko, Anna and Bridgland, Alex and Meyer, Clemens and Kohl, Simon A. A. and Ballard, Andrew J. and Cowie, Andrew and Romera-Paredes, Bernardino and Nikolov, Stanislav and Jain, Rishub and Adler, Jonas and Back, Trevor and Petersen, Stig and Reiman, David and Clancy, Ellen and Zielinski, Michal and Steinegger, Martin and Pacholska, Michalina and Berghammer, Tamas and Bodenstein, Sebastian and Silver, David and Vinyals, Oriol and Senior, Andrew W. and Kavukcuoglu, Koray and Kohli, Pushmeet and Hassabis, Demis},
  title   = {Highly accurate protein structure prediction with {AlphaFold}},
  journal = {Nature},
  volume  = {596},
  number  = {7873},
  pages   = {583--589},
  year    = {2021}
}

@article{lin2023evolutionary,
  title={Evolutionary-scale prediction of atomic-level protein structure with a language model},
  author={Lin, Zeming and Akin, Halil and Rao, Roshan and Hie, Brian and Zhu, Zhongkai and Lu, Wenting and Smetanin, Nikita and Verkuil, Robert and Kabeli, Ori and Shmueli, Yaniv and others},
  journal={Science},
  volume={379},
  number={6637},
  pages={1123--1130},
  year={2023},
  publisher={American Association for the Advancement of Science}
}

@article{abramson2024accurate,
  author  = {Abramson, Josh and Adler, Jonas and Dunger, Jack and Evans, Richard and Green, Tim and Pritzel, Alexander and Ronneberger, Olaf and Willmore, Lindsay and Ballard, Andrew J. and Bambrick, Jason and Bodenstein, Sebastian W. and Evans, David A. and Hung, Chia-Chun and O'Neill, Michael and Reiman, David and Tunyasuvunakool, Kathryn and Wu, Zachary and {\v{Z}}emgulyt{\.e}, Akvil{\.e} and Arvaniti, Eirini and Beattie, Charles and Bertolli, Ottavia and Bridgland, Alex and Cherepanov, Alexey and Congreve, Miles and Cowen-Rivers, Alexander I. and Cowie, Andrew and Figurnov, Michael and Fuchs, Fabian B. and Gladman, Hannah and Jain, Rishub and Khan, Yousuf A. and Low, Caroline M. R. and Perlin, Kuba and Potapenko, Anna and Savy, Pascal and Singh, Sukhdeep and Stecula, Adrian and Thillaisundaram, Ashok and Tong, Catherine and Yakneen, Sergei and Zhong, Ellen D. and Zielinski, Michal and {\v{Z}}{\'\i}dek, Augustin and Bapst, Victor and Kohli, Pushmeet and Jaderberg, Max and Hassabis, Demis and Jumper, John M.},
  title   = {Accurate structure prediction of biomolecular interactions with {AlphaFold} 3},
  journal = {Nature},
  volume  = {630},
  number  = {8016},
  pages   = {493--500},
  year    = {2024}
}

@article{watson2023de,
  author  = {Watson, Joseph L. and Juergens, David and Bennett, Nathaniel R. and Trippe, Brian L. and Yim, Jason and Eisenach, Helen E. and Ahern, Woody and Borst, Andrew J. and Ragotte, Robert J. and Milles, Lukas F. and Wicky, Basile I. M. and Hanikel, Nikita and Pellock, Samuel J. and Courbet, Alexis and Sheffler, William and Wang, Jue and Venkatesh, Preetham and Sappington, Isaac and Torres, Susana V{\'a}zquez and Lauko, Anna and {De Bortoli}, Valentin and Mathieu, Eric and Ovchinnikov, Sergey and Barzilay, Regina and Jaakkola, Tommi S. and DiMaio, Frank and Baek, Minkyung and Baker, David},
  title   = {De novo design of protein structure and function with {RFdiffusion}},
  journal = {Nature},
  volume  = {620},
  number  = {7976},
  pages   = {1089--1100},
  year    = {2023}
}

@article{ingraham2023illuminating,
  author  = {Ingraham, John B. and Baranov, Max and Costello, Zak and Barber, Karl W. and Wang, Wujie and Ismail, Ahmed and Frappier, Vincent and Lord, Dana M. and Ng-Thow-Hing, Christopher and {Van Vlack}, Emilien R. and Tie, Shan and Xue, Vincent and Cowles, Sarah C. and Leung, Alan and Rodrigues, Jo{\~a}o V. and Morales-Perez, Claudio L. and Ayoub, Alex M. and Green, Robin and Puentes, Katherine and Oplinger, Frank and Panwar, Nishant V. and Obermeyer, Fritz and Root, Andrew R. and Beam, Andrew L. and Poelwijk, Frank J. and Grigoryan, Gevorg},
  title   = {Illuminating protein space with a programmable generative model},
  journal = {Nature},
  volume  = {623},
  number  = {7989},
  pages   = {1070--1078},
  year    = {2023}
}

@article{hayes2025simulating,
  author  = {Hayes, Thomas and Rao, Roshan and Akin, Halil and Sofroniew, Nicholas J. and Oktay, Deniz and Lin, Zeming and Verkuil, Robert and Tran, Vincent Q. and Deaton, Jonathan and Wiggert, Marius and Badkundri, Rohil and Shafkat, Irhum and Gong, Jun and Derry, Alexander and Molina, Rafael S. and Thomas, Neil and Khan, Yousuf A. and Mishra, Chetan and Kim, Carolyn and Bartie, Liam J. and Nemeth, Marinka and Hsu, Patrick D. and Sercu, Tom and Candido, Salvatore and Rives, Alexander},
  title   = {Simulating 500 million years of evolution with a language model},
  journal = {Science},
  volume  = {387},
  number  = {6736},
  pages   = {850--858},
  year    = {2025}
}

@article{dauparas2022robust,
  author  = {Dauparas, Justas and Anishchenko, Ivan and Bennett, Nathaniel and Bai, Hua and Ragotte, Robert J. and Milles, Lukas F. and Wicky, Basile I. M. and Courbet, Alexis and de Haas, Robert J. and Bethel, Neville and Leung, Paul J. Y. and Huddy, Timothy F. and Pellock, Samuel and Tischer, Doug and Chan, F{\'a}tima and Koepnick, Brian and Nguyen, Hannah and Kang, Alex and Sankaran, Banumathi and Bera, Asim K. and King, Neil P. and Baker, David},
  title   = {Robust deep learning--based protein sequence design using {ProteinMPNN}},
  journal = {Science},
  volume  = {378},
  number  = {6615},
  pages   = {49--56},
  year    = {2022}
}

@article{stark2025boltzgen,
  author  = {Stark, Hannes and Faltings, Felix and Choi, MinGyu and Xie, Yuxin and Hur, Eunsu and O'Donnell, Timothy and Bushuiev, Anton and U{\c{c}}ar, Talip and Passaro, Saro and Mao, Weian and Reveiz, Mateo and Bushuiev, Roman and Pluskal, Tom{\'a}{\v{s}} and Sivic, Josef and Kreis, Karsten and Vahdat, Arash and Ray, Shamayeeta and Goldstein, Jonathan T. and Savinov, Andrew and Hambalek, Jacob A. and Gupta, Anshika and Taquiri-Diaz, Diego A. and Zhang, Yaotian and Hatstat, A. Katherine and Arada, Angelika and Kim, Nam Hyeong and Tackie-Yarboi, Ethel and Boselli, Dylan and Schnaider, Lee and Liu, Chang C. and Li, Gene-Wei and Hnisz, Denes and Sabatini, David M. and DeGrado, William F. and Wohlwend, Jeremy and Corso, Gabriele and Barzilay, Regina and Jaakkola, Tommi},
  title   = {{BoltzGen}: Toward universal binder design},
  journal = {bioRxiv},
  year    = {2025},
  doi     = {10.1101/2025.11.20.689494},
  note    = {Preprint}
}

@article{zhang2025odesign,
  author  = {Zhang, Odin and Zhang, Xujun and Lin, Haitao and Tan, Cheng and Wang, Qinghan and Mo, Yuanle and Feng, Qiantai and Du, Gang and Yu, Yuntao and Jin, Zichang and You, Ziyi and Lin, Peicong and Zhang, Yijie and Tao, Yuyang and Chen, Shicheng and Chen, Jack Xiaoyu and Hua, Chenqing and Zhao, Weibo and Ma, Runze and Xia, Yunpeng and Ying, Kejun and Li, Jun and Zeng, Yundian and Lang, Lijun and Pan, Peichen and Cao, Hanqun and Song, Zihao and Qiang, Bo and Wang, Jiaqi and Ji, Pengfei and Bai, Lei and Zhang, Jian and Hsieh, Chang-Yu and Heng, Pheng Ann and Sun, Siqi and Hou, Tingjun and Zheng, Shuangjia},
  title   = {{ODesign}: A world model for biomolecular interaction design},
  journal = {arXiv preprint arXiv:2510.22304},
  year    = {2025}
}

@inproceedings{liu2024a,
  author    = {Liu, Shengchao and Li, Yuxuan and Li, Zhuoxinran and Gitter, Anthony and Tang, Jian},
  title     = {A text-guided protein design framework},
  booktitle = {Advances in Neural Information Processing Systems},
  year      = {2023},
  note      = {ProteinDT}
}

@article{weng2025disprotedit,
  author  = {Weng, Yunjia and others},
  title   = {{DisProtEdit}: Improving diffusion-based protein generation via disentangled textual guidance},
  journal = {arXiv preprint},
  year    = {2025},
  note    = {Bibliographic details to be finalized}
}

@article{wang2024instructprotein,
  author  = {Wang, Zeyuan and others},
  title   = {{InstructProtein}: Aligning human and protein language via knowledge instruction},
  journal = {arXiv preprint},
  year    = {2024}
}

@article{lv2024prollama,
  author  = {Lv, Liuzhenghao and others},
  title   = {{ProLLaMA}: A protein large language model for multi-task protein language processing},
  journal = {arXiv preprint},
  year    = {2024}
}

@inproceedings{rao2019evaluating,
  author    = {Rao, Roshan and Bhattacharya, Nicholas and Thomas, Neil and Duan, Yan and Chen, Xi and Canny, John and Abbeel, Pieter and Song, Yun S.},
  title     = {Evaluating protein transfer learning with {TAPE}},
  booktitle = {Advances in Neural Information Processing Systems},
  year      = {2019}
}

@inproceedings{xu2022peer,
  author    = {Xu, Minghao and Zhang, Zuobai and Lu, Jiarui and Zhu, Zhaocheng and Zhang, Yangtian and Ma, Chang and Liu, Runcheng and Tang, Jian},
  title     = {{PEER}: A comprehensive and multi-task benchmark for protein sequence understanding},
  booktitle = {Advances in Neural Information Processing Systems},
  year      = {2022}
}

@inproceedings{notin2023proteingym,
  author    = {Notin, Pascal and Kollasch, Aaron and Ritter, Daniel and {van Niekerk}, Lood and Paul, Steffanie and Spinner, Han and Rollins, Nathan and Shaw, Ada and Wektoro, Rose and Frazer, Jonathan and Gal, Yarin and Marks, Debora},
  title     = {{ProteinGym}: Large-scale benchmarks for protein fitness prediction and design},
  booktitle = {Advances in Neural Information Processing Systems},
  year      = {2023}
}

@inproceedings{gao2023proteininvbench,
  author    = {Gao, Zhangyang and Tan, Cheng and Li, Stan Z.},
  title     = {{ProteinInvBench}: Benchmarking protein inverse folding on diverse tasks, models, and metrics},
  booktitle = {Advances in Neural Information Processing Systems},
  year      = {2023}
}

@inproceedings{chen2025scienceagentbench,
  author    = {Chen, Ziru and Chen, Shijie and Ning, Yuting and Zhang, Qianheng and Wang, Boshi and Yu, Botao and Li, Yifei and Liao, Zeyi and Wei, Chen and Lu, Ziao and others},
  title     = {{ScienceAgentBench}: Toward rigorous assessment of language agents for data-driven scientific discovery},
  booktitle = {International Conference on Learning Representations},
  year      = {2025}
}

@article{bran2024augmenting,
  author  = {Bran, Andres M. and Cox, Sam and Schilter, Oliver and Baldassari, Carlo and White, Andrew D. and Schwaller, Philippe},
  title   = {Augmenting large language models with chemistry tools},
  journal = {Nature Machine Intelligence},
  volume  = {6},
  pages   = {525--535},
  year    = {2024},
  note    = {ChemCrow}
}

@article{ghafarollahi2024protagents,
  author  = {Ghafarollahi, Alireza and Buehler, Markus J.},
  title   = {{ProtAgents}: Protein discovery via large language model multi-agent collaborations combining physics and machine learning},
  journal = {Digital Discovery},
  year    = {2024}
}

@article{kim2026biodesignbench,
  author  = {Kim, S. and Romero, P. A.},
  title   = {{BioDesignBench}: Evaluating language agents on expert-curated biomolecular design workflows},
  journal = {arXiv preprint},
  year    = {2026},
  note    = {Bibliographic details to be finalized}
}

@inproceedings{Rao2019,
  title     = {Evaluating Protein Transfer Learning with {TAPE}},
  author    = {Rao, Roshan and Bhattacharya, Nicholas and Thomas, Neil
               and Duan, Yan and Chen, Peter and Canny, John
               and Abbeel, Pieter and Song, Yun S.},
  booktitle = {Advances in Neural Information Processing Systems},
  volume    = {32},
  pages     = {9689--9701},
  year      = {2019},
  url       = {https://proceedings.neurips.cc/paper/2019/hash/37f65c068b7723cd7809ee2d31d7861c-Abstract.html}
}

@inproceedings{Notin2023,
  title     = {{ProteinGym}: Large-Scale Benchmarks for Protein Fitness Prediction and Design},
  author    = {Notin, Pascal and Kollasch, Aaron W. and Ritter, Daniel
               and van Niekerk, Lood and Paul, Steffanie and Spinner, Hansen
               and Rollins, Nathan and Shaw, Ada and Orenbuch, Rose
               and Weitzman, Ruben and Frazer, Jonathan and Dias, Mafalda
               and Franceschi, Dinko and Gal, Yarin and Marks, Debora S.},
  booktitle = {Advances in Neural Information Processing Systems},
  volume    = {36},
  pages     = {64331--64379},
  year      = {2023},
  doi       = {10.52202/075280-2810},
  url       = {https://proceedings.neurips.cc/paper_files/paper/2023/hash/cac723e5ff29f65e3fcbb0739ae91bee-Abstract-Datasets_and_Benchmarks.html}
}

@article{Liu2025,
  title   = {A Text-Guided Protein Design Framework},
  author  = {Liu, Shengchao and Li, Yanjing and Li, Zhuoxinran
             and Gitter, Anthony and Zhu, Yutao and Lu, Jiarui
             and Xu, Zhao and Nie, Weili and Ramanathan, Arvind
             and Xiao, Chaowei and Tang, Jian and Guo, Hongyu
             and Anandkumar, Anima},
  journal = {Nature Machine Intelligence},
  volume  = {7},
  pages   = {580--591},
  year    = {2025},
  doi     = {10.1038/s42256-025-01011-z}
}

@article{Watson2023,
  title   = {De Novo Design of Protein Structure and Function with {RFdiffusion}},
  author  = {Watson, Joseph L. and Juergens, David and Bennett, Nathaniel R.
             and Trippe, Brian L. and Yim, Jason and Eisenach, Helen E.
             and Ahern, Woody and Borst, Andrew J. and Ragotte, Robert J.
             and Milles, Lukas F. and Wicky, Basile I. M. and Hanikel, Nikita
             and Pellock, Samuel J. and Courbet, Alexis and Sheffler, William
             and Wang, Jue and Venkatesh, Preetham and Sappington, Isaac
             and V{\'a}zquez Torres, Susana and Lauko, Anna
             and De Bortoli, Valentin and Mathieu, Emile
             and Ovchinnikov, Sergey and Barzilay, Regina
             and Jaakkola, Tommi S. and DiMaio, Frank
             and Baek, Minkyung and Baker, David},
  journal = {Nature},
  volume  = {620},
  number  = {7976},
  pages   = {1089--1100},
  year    = {2023},
  doi     = {10.1038/s41586-023-06415-8}
}

@article{KabschSander1983,
  title   = {Dictionary of Protein Secondary Structure:
             Pattern Recognition of Hydrogen-Bonded and Geometrical Features},
  author  = {Kabsch, Wolfgang and Sander, Christian},
  journal = {Biopolymers},
  volume  = {22},
  number  = {12},
  pages   = {2577--2637},
  year    = {1983},
  doi     = {10.1002/bip.360221211}
}

@article{Touw2015,
  title   = {A Series of {PDB}-Related Databanks for Everyday Needs},
  author  = {Touw, Wouter G. and Baakman, Coos and Black, Jon
             and te Beek, Tim A. H. and Krieger, Elmar
             and Joosten, Robbie P. and Vriend, Gert},
  journal = {Nucleic Acids Research},
  volume  = {43},
  number  = {D1},
  pages   = {D364--D368},
  year    = {2015},
  doi     = {10.1093/nar/gku1028}
}

@article{ChangeuxEdelstein2005,
  title   = {Allosteric Mechanisms of Signal Transduction},
  author  = {Changeux, Jean-Pierre and Edelstein, Stuart J.},
  journal = {Science},
  volume  = {308},
  number  = {5727},
  pages   = {1424--1428},
  year    = {2005},
  doi     = {10.1126/science.1108595}
}

@article{Jones1986,
  title   = {Replacing the Complementarity-Determining Regions in a Human Antibody with Those from a Mouse},
  author  = {Jones, Peter T. and Dear, Paul H. and Foote, Jefferson
             and Neuberger, Michael S. and Winter, Greg},
  journal = {Nature},
  volume  = {321},
  number  = {6069},
  pages   = {522--525},
  year    = {1986},
  doi     = {10.1038/321522a0}
}

@article{Ofek2010,
  title   = {Elicitation of Structure-Specific Antibodies by Epitope Scaffolds},
  author  = {Ofek, Gilad and Guenaga, F. Javier and Schief, William R.
             and Skinner, Jeff and Baker, David and Wyatt, Richard
             and Kwong, Peter D.},
  journal = {Proceedings of the National Academy of Sciences},
  volume  = {107},
  number  = {42},
  pages   = {17880--17887},
  year    = {2010},
  doi     = {10.1073/pnas.1004728107}
}

@article{Tien2013,
  title   = {Maximum Allowed Solvent Accessibilites of Residues in Proteins},
  author  = {Tien, Matthew Z. and Meyer, Austin G. and Sydykova, Dariya K.
             and Spielman, Stephanie J. and Wilke, Claus O.},
  journal = {PLOS ONE},
  volume  = {8},
  number  = {11},
  pages   = {e80635},
  year    = {2013},
  doi     = {10.1371/journal.pone.0080635}
}

@article{Chothia1984,
  title   = {Principles That Determine the Structure of Proteins},
  author  = {Chothia, Cyrus},
  journal = {Annual Review of Biochemistry},
  volume  = {53},
  pages   = {537--572},
  year    = {1984},
  doi     = {10.1146/annurev.bi.53.070184.002541}
}

@article{Rose1985,
  title   = {Hydrophobicity of Amino Acid Residues in Globular Proteins},
  author  = {Rose, George D. and Geselowitz, Allan R.
             and Lesser, George J. and Lee, Richard H.
             and Zehfus, Mark H.},
  journal = {Science},
  volume  = {229},
  number  = {4716},
  pages   = {834--838},
  year    = {1985},
  doi     = {10.1126/science.4023714}
}

@article{Rettie2025a,
  title   = {Cyclic Peptide Structure Prediction and Design Using {AlphaFold2}},
  author  = {Rettie, Stephen A. and Campbell, Katelyn V.
             and Bera, Asim K. and Kang, Alex and Kozlov, Simon
             and Flores Bueso, Yensi and De La Cruz, Joshmyn
             and Ahlrichs, Maggie and Cheng, Suna and Gerben, Stacey R.
             and Lamb, Mila and Murray, Analisa and Adebomi, Victor
             and Zhou, Guangfeng and DiMaio, Frank
             and Ovchinnikov, Sergey and Bhardwaj, Gaurav},
  journal = {Nature Communications},
  volume  = {16},
  pages   = {4730},
  year    = {2025},
  doi     = {10.1038/s41467-025-59940-7}
}

@article{Rettie2025b,
  title   = {Accurate De Novo Design of High-Affinity Protein-Binding Macrocycles Using Deep Learning},
  author  = {Rettie, Stephen A. and Juergens, David and Adebomi, Victor
             and Flores Bueso, Yensi and Zhao, Qinqin
             and Leveille, Alexandria N. and Liu, Andi and Bera, Asim K.
             and Wilms, Joana A. and {\"U}ffing, Alina and Kang, Alex
             and Brackenbrough, Evans and Lamb, Mila and Gerben, Stacey R.
             and Murray, Analisa and Levine, Paul M. and Schneider, Maika
             and Vasireddy, Vibha and Ovchinnikov, Sergey
             and Weiergr{\"a}ber, Oliver H. and Willbold, Dieter
             and Kritzer, Joshua A. and Mougous, Joseph D.
             and Baker, David and DiMaio, Frank and Bhardwaj, Gaurav},
  journal = {Nature Chemical Biology},
  volume  = {21},
  number  = {12},
  pages   = {1948--1956},
  year    = {2025},
  doi     = {10.1038/s41589-025-01929-w}
}

@misc{Zheng2025,
  title         = {{MotifBench}: A Standardized Protein Design Benchmark for Motif-Scaffolding Problems},
  author        = {Zheng, Zhuoqi and Zhang, Bo and Didi, Kieran
                   and Yang, Kevin K. and Yim, Jason
                   and Watson, Joseph L. and Chen, Hai-Feng
                   and Trippe, Brian L.},
  year          = {2025},
  eprint        = {2502.12479},
  archivePrefix = {arXiv},
  primaryClass  = {q-bio.BM},
  doi           = {10.48550/arXiv.2502.12479}
}

@article{Dauparas2022,
  title   = {Robust Deep Learning-Based Protein Sequence Design Using {ProteinMPNN}},
  author  = {Dauparas, Justas and Anishchenko, Ivan and Bennett, Nathaniel
             and Bai, Hua and Ragotte, Robert J. and Milles, Lukas F.
             and Wicky, Basile I. M. and Courbet, Alexis
             and de Haas, Rob J. and Bethel, Nathaniel
             and Leung, Po-Ssu Jason Y. and Huddy, Timothy F.
             and Pellock, Samuel and Tischer, Doug and Chan, Frederick
             and Koepnick, Brian and Nguyen, Huey and Kang, Alex
             and Sankaran, Banumathi and Baker, David and others},
  journal = {Science},
  volume  = {378},
  number  = {6615},
  pages   = {49--56},
  year    = {2022},
  doi     = {10.1126/science.add2187}
}

@article{Lin2023,
  title   = {Evolutionary-Scale Prediction of Atomic-Level Protein Structure with a Language Model},
  author  = {Lin, Zeming and Akin, Halil and Rao, Roshan
             and Hie, Brian and Zhu, Zhongkai and Lu, Wenting
             and Smetanin, Nikita and Verkuil, Robert and Kabeli, Ori
             and Shmueli, Yaniv and dos Santos Costa, Allan
             and Fazel-Zarandi, Maryam and Sercu, Tom
             and Candido, Salvatore and Rives, Alexander},
  journal = {Science},
  volume  = {379},
  number  = {6637},
  pages   = {1123--1130},
  year    = {2023},
  doi     = {10.1126/science.ade2574}
}

@article{Berman2000,
  author  = {Berman, Helen M. and Westbrook, John and Feng, Zukang
             and Gilliland, Gary and Bhat, T. N. and Weissig, Helge
             and Shindyalov, Ilya N. and Bourne, Philip E.},
  title   = {The {Protein Data Bank}},
  journal = {Nucleic Acids Research},
  volume  = {28},
  number  = {1},
  pages   = {235--242},
  year    = {2000},
  doi     = {10.1093/nar/28.1.235}
}

@article{butcher2025rfd3,
  title   = {De Novo Design of All-Atom Biomolecular Interactions with {RFdiffusion3}},
  author  = {Butcher, Nathan and Watson, Joseph L. and Bennett, Nathaniel R. and Lee, Jueun and Sumida, Brian and Yang, Frank and Coventry, Brian and Baker, David},
  journal = {bioRxiv preprint},
  doi     = {10.1101/2025.09.18.676967},
  year    = {2025}
}

@misc{anthropic2025claude,
  title        = {The {Claude} 3.5 and Beyond Model Family},
  author       = {{Anthropic}},
  year         = {2025},
  howpublished = {\url{https://www.anthropic.com/claude}}
}

@misc{openai2025gpt,
  title        = {{GPT-4} and Frontier Generative Architecture Series},
  author       = {{OpenAI}},
  year         = {2025},
  howpublished = {\url{https://openai.com/research}}
}

@article{deepseek2025deepseek,
  title   = {{DeepSeek-V3} Technical Report},
  author  = {Liu, Aixin and Feng, Bei and Wang, Bin and Wang, Bingxuan and Liu, Bo and Chen, Chenggang and Deng, Chengqi and Chen, Chong and Zhao, Chunjing and Liang, Wenfeng and others},
  journal = {arXiv preprint arXiv:2412.19437},
  year    = {2024}
}

@article{qwen2025qwen,
  title   = {{Qwen2.5} Technical Report},
  author  = {Yang, An and Yang, Baosong and Zhang, Binyuan and Hui, Bofei and Zheng, Bo and Yu, Bowen and Li, Dayi and Liu, Dayiheng and Huang, Fei and Wei, Haoran and others},
  journal = {arXiv preprint arXiv:2412.15115},
  year    = {2024}
}

@misc{moonshot2025kimi,
  title        = {{Kimi} and {Moonshot} Architecture Series},
  author       = {{Moonshot AI}},
  year         = {2025},
  howpublished = {\url{https://www.moonshot.ai}}
}

@article{zheng2025motifbench,
  title   = {{MotifBench}: A Comprehensive Benchmark for Functional Motif Scaffolding},
  author  = {Zheng, Zaixiang and others},
  journal = {bioRxiv preprint},
  year    = {2025}
}
\bibliographystyle{iclr2027_conference}

\newpage
\appendix
%
\section{eight PROTEAN edit types.}
See Table \ref{tab:edit-types}. We choose some hyperparameters such as 0.35, 0.10 because of the corresponding statistical quantile.
\begin{table*}[t]
\centering
\caption{
The eight PROTEAN edit types.
Thresholds are frozen before model evaluation.
The final column gives the biological motivation for each structural intervention rather than an experimentally validated outcome.
}
\label{tab:edit-types}
\begingroup
\footnotesize
\setlength{\tabcolsep}{3pt}
\renewcommand{\arraystretch}{0.96}
\begin{tabular}{@{}p{0.14\linewidth} c p{0.13\linewidth} p{0.31\linewidth} p{0.26\linewidth}@{}}

\toprule
\textbf{Edit type} &
\textbf{Sub-tasks} &
\textbf{Additional input} &
\textbf{Primary criterion} &
\textbf{Biological motivation} \\
\midrule

Helix enrichment &
50 &
-- &
secondary structure propensity contrast $\ge 0.35$ &
local rigidification and presentation of interaction surfaces \\

Sheet enrichment &
50 &
-- &
secondary structure propensity contrast $\ge 0.10$ &
extended recognition geometries and strand-rich structural states \\

Inter-domain approach &
50 &
reference patch &
source-referenced median distance decreases by at least $1.0$~\AA{} &
domain closure, interface formation, and allosteric rearrangement \\

Inter-domain separation &
50 &
reference patch &
source-referenced median distance increases by at least $1.0$~\AA{} &
domain opening, relief of autoinhibition, and changes in multivalent geometry \\

Reference-fragment mimicry &
50 &
reference fragment and residue map &
backbone RMSD $\le 2.0$~\AA{}, at least 3 mapped residues, coverage $\ge 0.80$ &
transfer of binding, catalytic, or epitope geometries \\

Core burial &
50 &
-- &
median RSA decreases by at least $0.10$, final RSA $\le 0.40$, with constrained local compaction &
packing, solvent shielding, and stabilization-oriented redesign \\

Head-to-tail cyclization &
50 &
terminal editable region &
one new head-to-tail amide bond with N--C distance in $[1.20,1.50]$~\AA{} &
conformational constraint and proteolytic stabilization of peptides \\

Motif scaffolding &
30 &
conserved motif segments &
motif preservation plus the MotifBench refolding criterion &
transfer of a functional geometry onto a new structural carrier \\

\bottomrule
\end{tabular}

\endgroup

\end{table*}
\section{Worked Examples and Task Construction}
\label{app:examples}

This appendix grounds the benchmark definition in concrete artifacts: how the
tasks were constructed (Section~\ref{app:pipeline}), what a task card contains
for each of the eight edit types (Section~\ref{app:cards}), the exact prompt a
model receives and the specification it returned (Sections~\ref{app:prompt}
and~\ref{app:rollouts}), rendered source/candidate structures for two
successful edits and one cyclization rollout that closes the ring yet
fails the scope tolerance (Section~\ref{app:figures}),
pseudocode for the three gates (Section~\ref{app:gatepseudo}), and detailed
bottleneck ablations tracking structural leakage and biophysical refolding attrition (Section~\ref{app:bottlenecks}).

\subsection{Task construction pipeline}
\label{app:pipeline}

\begin{figure*}[t]
\centering
\begin{tikzpicture}[
  box/.style={draw, rounded corners, align=center, font=\tiny, inner sep=2.5pt},
  stage/.style={box, fill=gray!12, text width=13.0cm},
  fam/.style={box, fill=white, text width=3.75cm},
  emph/.style={box, fill=red!8, very thick},
  import/.style={box, fill=green!10, dashed, text width=5.2cm},
  arr/.style={->, thick}]
\node[stage] (acq) at (0,0) {\textbf{1. Source acquisition}\\
  experimental PDB leg: UniProt cross-refs ($\le$ 3 entries/accession, X-ray/EM), quality filters from the
  interface-loop harvest (X-ray $\le$ 3.0\,\AA, $R_{\mathrm{free}}\!$$\le\!$0.30, chains 40--300/80--700\,aa, BSA $500$--$2500$\,\AA$^2$)\\
  $\parallel$ \ AFDB leg: Swiss-Prot reviewed (human), length 100--800, coverage $\ge$ 0.90, pLDDT $\ge$ 70};
\node[stage, emph] (iso) at (0,-1.7) {\textbf{2. Source isolation, before construction:} similarity graph over the source pool ---
  MMseqs2 at $\ge$30\% identity with $\ge$70\% mutual coverage ($\ge$80\% one-sided)\\
  $+$ five hard edge classes (exact duplicates, shared accessions, superfamilies, assemblies, shared interfaces)
  $\rightarrow$ connected components; every task inherits its component};
\draw[rounded corners, draw=black!55, fill=orange!10] (-6.7,-2.55) rectangle (6.7,-5.75);
\node[font=\tiny\bfseries, align=center, text width=12.6cm] at (0,-3.0) {3. Family construction --- six families in parallel, yielding the seven equal-length task types (200 tasks each)};
\node[fam] (move)  at (-4.25,-3.95) {\textbf{move} (approach / separation)\\ interface loops, $6$--$12$\,\AA\ from landmark};
\node[fam] (ss)    at (0,-3.95)    {\textbf{helix / sheet enrichment}\\ paired-structure witness};
\node[fam] (rfm)   at (4.25,-3.95) {\textbf{fragment mimicry}\\ same-protein PDB pair, $\le 2$\,\AA};
\node[fam] (shape) at (-4.25,-5.1) {\textbf{shape compaction}\\ compaction witness, centred window};
\node[fam] (cyc)   at (0,-5.1)    {\textbf{cyclization}\\ opened real cyclic peptides};
\node[fam] (bur)   at (4.25,-5.1) {\textbf{burial} (v3)\\ 1:1 from shape, exposed windows};
\node[stage] (quota) at (0,-6.3) {\textbf{4. Quota sampling:} $\le$ 8 tasks per source component, split bucket fixed by the component before ranking};
\node[stage] (can) at (0,-7.1) {\textbf{5. Canonical view:} chains renamed A, B, \dots{}, residues renumbered $1..N$ (318 canonical sources)};
\node[stage, text width=6.8cm] (freeze) at (0,-8.3) {\textbf{6. Freeze and audit:} manifest frozen before evaluation;
  first-manifest audit found 13 concealed shared sources $\rightarrow$ rebuilt with indivisible source entries;
  the complete suite is the evaluation set, with no selection};
\node[import] (motif) at (0,-10.0) {\textbf{motif scaffolding}\\ imported directly from MotifBench
  (30 test cases); segments split per chain, balanced-deterministic gaps; own output contract ---
  does not pass through the source-pool lineage};
\draw[arr] (acq.south) -- (iso.north);
\draw[arr] (iso.south) -- (0,-2.55);
\draw[arr] (0,-5.75) -- (quota.north);
\draw[arr] (quota.south) -- (can.north);
\draw[arr] (can.south) -- (freeze.north);
\draw[arr] (motif.north) -- (freeze.south);
\end{tikzpicture}
\caption{Task-construction pipeline. Sources come from two filtered legs, and
\emph{each source is grouped at intake}. The load-bearing step is the
source-isolation graph, computed on the source pool \emph{before} any task is
constructed: near-duplicates and related structures are merged into connected
components, and every task inherits its component. The seven equal-length
task types are then built in parallel from the grouped pool---six
construction families, with the move and secondary-structure families each
yielding two task types, and burial derived one-to-one from the shape
family. Motif scaffolding is \emph{imported directly from MotifBench} under
its own output contract; it does not pass through the source-acquisition or
isolation lineage and joins the suite only at the frozen evaluation set.
Quota sampling caps tasks per component ($\le 8$, split bucket fixed by the
component), the canonical view renumbers structures for model input, and the
manifest is frozen after audit---the complete suite (350 
$+$ 30 motif) forms the evaluation set.}
\label{fig:pipeline}
\end{figure*}
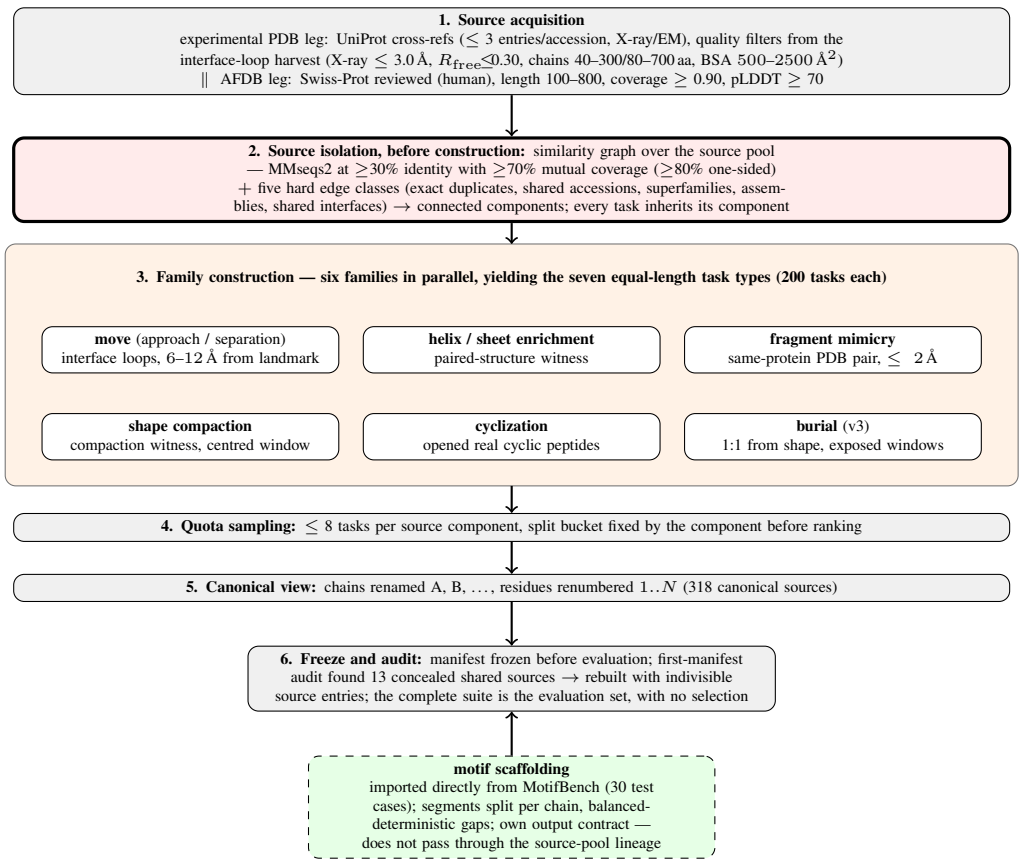

Figure~\ref{fig:pipeline} summarizes how the evaluation set was produced.
Parameters quoted on the edges are the frozen construction values; all are
recorded in the construction scripts and build manifests.

\subsubsection{Sources.}\label{app:pipelinesource}
The experimental leg collects cross-referenced PDB entries for reviewed
UniProt accessions (at most three entries per accession, X-ray and
electron-microscopy preferred) and inherits the quality filters of the
interface-loop harvest that seeded the editable-region geometry:
X-ray resolution $\le 3.0$\,\AA{}, $R_{\mathrm{free}} \le 0.30$ and
$R_{\mathrm{free}} - R_{\mathrm{work}} \le 0.08$, editable chain 40--300
residues and fixed chains 80--700 residues, total modelled length
$\le 900$ residues, buried surface area 500--2{,}500\,\AA$^2$, and at least
eight interface residues on each side. The predicted leg uses reviewed human
Swiss-Prot entries of length 100--800 whose AlphaFold Database model has
sequence coverage $\ge 0.90$ and pLDDT $\ge 70$ both globally and over the
edit region. The release mixes 1{,}142 experimental and 258 predicted tasks
across the seven equal-length families.

\subsubsection{Per-family construction.}
Editable windows are selected under family-specific feasibility conditions:
distance tasks require a source-to-patch median distance in 6--12\,\AA{}
(centred at 9\,\AA{} by the selection score); secondary-structure tasks pair
the source with another experimental structure of the same protein whose helix
(or strand) count exceeds the source's by at least three over the window;
mimicry tasks align a same-accession PDB pair globally and take 24--48-residue
blocks whose backbone Kabsch RMSD is at most 2\,\AA{} (the residue map is the
aligned block); shape tasks use a paired witness with a 5\% smaller radius of
gyration and rebuild a centred window of width $\min(80, \max(40, n/3))$;
cyclization tasks are \emph{opened} real cyclic peptides---nine witness chains
of 13--40 residues whose termini are rotated in $5^\circ$ steps until the
N--C distance lands in 2.0--4.0\,\AA{}, with the first and last four residues
forming the editable region.

\subsubsection{Motif import.}\label{app:motif}
The 30 motif-scaffolding cases are MotifBench test cases: the multi-segment
motif is split into per-segment structures, and scaffold gaps distribute the
remaining length deterministically over segment slots
(case 1: $52 + 21 + 52 = 125$ residues). These canonical gaps are
construction bookkeeping only: the model-facing instruction leaves the
scaffold lengths open and requires only arithmetic self-consistency with the
public total length (Table~\ref{tab:card-instances}).

\subsubsection{Canonical view, quotas, isolation, freeze.}\label{app:pipelineblind}
For model input and scoring, every source is canonicalized (chains renamed
A, B, \dots{}, residues renumbered $1..N$; 318 canonical sources). Quota
sampling yields 200 tasks per family with at most eight tasks per source
group. The source-isolation graph (MMseqs2 at $\ge$30\% identity with
$\ge$70\% mutual coverage, $\ge$80\% one-sided for short sequences, plus the
five hard edge classes of Appendix A) groups related sources; the manifest was
frozen before evaluation after an audit found thirteen task groupings that
concealed a shared source. The final evaluation set is the complete suite with
no held-out selection.

\subsection{Task cards, one per edit type}
\label{app:cards}

Each task ships as a card carrying the instruction (verbatim, English and
Chinese), the role partition as per-chain residue ranges, the sequence policy,
and the public target criteria. Table~\ref{tab:card-instances} gives one real
card per edit type (compact excerpt; long index arrays elided). The card
received by a model is the same object minus all routing metadata.

\begin{table*}[t]
\centering
\caption{One real task card per edit type (excerpt). Ranges are
\texttt{chain:label-seq}; the editable region is listed under ``edit scope''
with the reference landmark where applicable.}
\label{tab:card-instances}
\small
\setlength{\tabcolsep}{4pt}
\begin{tabular}{p{0.23\linewidth} p{0.30\linewidth} p{0.37\linewidth}}
\toprule
\textbf{Type / task id} & \textbf{Edit scope (chain:residues)} & \textbf{Instruction (English, verbatim)} \\
\midrule
Helix enrichment \\
\texttt{ptask-0227c7e0\ldots} &
\texttt{D:313-360} &
``Make the red edit core alpha-rich; substitutions only inside the core and preserve protected sequences.'' \\
\midrule
Sheet enrichment \\
\texttt{ptask-00201762\ldots} &
\texttt{A:97-144} &
``Make the red edit core beta-rich; substitutions only inside the core and preserve protected sequences.'' \\
\midrule
Inter-domain approach \\
\texttt{ptask-1c11ae98\ldots} &
\texttt{B:99-110}; ref.\ \texttt{A:7-12,193,196-197,201} &
``Move the red edit core closer to the purple reference; substitutions only inside the edit core.'' \\
\midrule
Inter-domain separation \\
\texttt{ptask-00a70700\ldots} &
\texttt{A:113-118}; ref.\ \texttt{B:29,111,114-116} &
``Move the red edit core away from the purple reference; substitutions only inside the edit core.'' \\
\midrule
Reference-fragment mimicry \\
\texttt{ptask-01425f04\ldots} &
\texttt{A:49-96}; ref.\ \texttt{2w8b\_A} + residue map &
``Make the red target fragment mimic the purple reference backbone; substitutions only inside the core.'' \\
\midrule
Core burial \\
\texttt{burial\_\allowbreak psource-3c1d\ldots} &
\texttt{A} (12-res.\ window) &
``Bury the red edit core: reduce its solvent accessibility (median RSA drop) while compacting it; substitutions only inside the edit core.'' \\
\midrule
Head-to-tail cyclization \\
\texttt{ptask-030bf23a\ldots} &
\texttt{A:1-4,13-16} &
``Cyclize the red N/C-terminal edit core with one declared head-to-tail amide bond; substitutions only in the core.'' \\
\midrule
Motif scaffolding \\
\texttt{motif\_case\_12} &
motifs \texttt{M0} (7 res.) $+$ \texttt{M1} (17 res.); redesign \texttt{M0}:1,7, \texttt{M1}:1,3,8,16,17 &
``Scaffold the fixed motifs M0 (7 residues) and M1 (17 residues) into a single chain of 100 residues, choosing the scaffold segment lengths yourself; the motif backbone geometry must remain fixed and residues 1, 7 of M0 and residues 1, 3, 8, 16, 17 of M1 may undergo sequence redesign.'' \\
\bottomrule
\end{tabular}
\end{table*}

The full card of the inter-domain-approach example used throughout this
appendix:
\begin{quote}\scriptsize
\begin{verbatim}
{"task_id": "ptask-1c11ae980dba27dd3a12", "action": "move_closer",
 "instruction": {"en": "Move the red edit core closer to the purple
   reference; substitutions only inside the edit core.", "zh": "..."},
 "roles": {
   "edit_core":        {"B": {"label_seq_ranges": "99-110"}},
   "relax_shell":      {"B": {"label_seq_ranges": "91-98,111-118"}},
   "hard_protected":   {"B": {"label_seq_ranges": "1-90,119-128"}},
   "fixed_context":    {"A": {"label_seq_ranges": "1-202"}},
   "reference_patch":  {"A": {"label_seq_ranges": "7-12,193,196-197,201"}}},
 "sequence_policy": {"profile": "core_redesign_v1", "mode": "core_redesign",
   "editable_regions": ["edit_core"],
   "protected_regions": ["relax_shell", "reference_patch",
                         "fixed_context", "hard_protected"],
   "max_substitution_fraction": null, "preserve_length": true,
   "preserve_chain_ids": true, "preserve_residue_order": true,
   "preserve_atom_mapping": true, "forbid_noncanonical": true},
 "target_criteria": {"primary_metric_id": "median_min_pseudo_cb_distance_delta_v2",
   "target": {"metric_id": "median_min_pseudo_cb_distance_delta_v2",
     "move_away":  {"delta_threshold_angstrom": 0.5},
     "move_closer": {"delta_threshold_angstrom": -0.5},
     "source_distance_angstrom": 8.954}},
 "source": {...}, "split": "test", "benchmark_split": "test",
 "source_content_group_id": "isolation-0daaf9b03d0388fabc69"}
\end{verbatim}
\end{quote}
The card carries the thresholds of the profile that constructed it; under the
frozen evaluation profile the move gate is $\pm 1.0$\,\AA{} and the
card-level value is ignored (the scorer flags such rows).

\subsection{What the model sees}
\label{app:prompt}

Each attempt is a single API call with two messages. The system prompt is
fixed per backend:
\begin{quote}\small\ttfamily
You are a protein design engineer. Translate the benchmark task into a
\{backend\} input specification. Respond with a single JSON object and nothing
else.
\end{quote}
The user message concatenates four segments (12{,}462 characters for the
worked example):
\begin{quote}\scriptsize
\begin{verbatim}
# Backend Capability Card
<markdown card: input schema, contig/design syntax, hotspot and bond
 channels, expressivity boundary, legality rules>

# Scoring Criteria
<markdown card rendered from the frozen scorer constants>

# Task Card
<the task-card JSON above, with all routing metadata removed:
 routing, backend_hint, required_primitives, backend_support,
 allowed_backends, preferred_backend, support_reason>

# Feedback on your previous attempt        (round >= 2, on failure only)
<public gate outcomes and failure classes>

Output the JSON input specification now.
\end{verbatim}
\end{quote}
Sampling is deterministic given the seed (temperature 0.2); at most one
structured retry is issued for malformed output. Because the protocol forbids
prose, the model's response \emph{is} the final backend specification---there
is no separate reasoning trace to display. Every request and response is
logged verbatim per task and round, and the examples below quote those logs.

\subsection{Rollouts: prompt, specification, and outcome}
\label{app:rollouts}

\paragraph{Success (inter-domain approach, DeepSeek $\times$ ODesign).}
For the worked card above, the model returned the ODesign specification
\begin{quote}\scriptsize
\begin{verbatim}
{"chains": [{"chain_type": "proteinChain", "sequence": "A/1-202", "if_cyc": false},
            {"chain_type": "proteinChain", "sequence": "B/1-98,12-12,B/111-128",
             "if_cyc": false}],
 "hotspot": "A/7,A/8,A/9,A/10,A/11,A/12,A/193,A/196,A/197,A/201",
 "partial_diff": "B/99-110"}
\end{verbatim}
\end{quote}
The specification is worth reading closely. Chain A (the fixed context) is
kept in full; on chain B the model kept residues 1--98 and 111--128 and asked
for an anonymous twelve-residue redesign segment (the \texttt{12-12} contig,
matching the editable region's length) whose position makes it cover exactly
\texttt{B/99-110}; the diffusion region \texttt{partial\_diff} names the same
span; and the ten hotspots are precisely the reference-patch residues listed
on the card. In other words, the model derived the keep/redesign split, the
diffusion window, and the landmark conditioning from the task card---the
compilation skill the benchmark measures.

\paragraph{Success (helix enrichment, DeepSeek $\times$ ODesign).}
For the card whose edit core is \texttt{A:97-144} (48 residues), the model
returned the ODesign specification
\begin{quote}\scriptsize
\begin{verbatim}
{"chains": [{"chain_type": "proteinChain",
             "sequence": "A/1-96,48-48,A/145-282", "if_cyc": false}],
 "hotspot": "", "partial_diff": "A/97-144"}
\end{verbatim}
\end{quote}
The anonymous $48$-residue redesign contig (\texttt{48-48}) is positioned so
that it covers exactly \texttt{A/97-144}, and the diffusion region names the
same span---the same keep/redesign compilation skill as the approach
example, at six times the region size. The candidate's editable region,
coil throughout in the source, is remodelled into helices covering $56\%$
of the window (Figure~\ref{fig:alpha-example}).

\paragraph{Cyclization: target pass, scope fail (DeepSeek $\times$
BoltzGen).}
For a $22$-residue opened cyclic peptide (edit core \texttt{A:1-4,19-22}),
the model returned the BoltzGen specification
\begin{quote}\scriptsize
\begin{verbatim}
{"entities": [{"file": {"include": [{"chain": {"id": "A"}}],
                       "design": [{"chain": {"id": "A", "res_index": "1..4"}},
                                   {"chain": {"id": "A", "res_index": "19..22"}}]}}],
 "constraints": [{"bond": {"atom1": ["A", 22, "C"],
                           "atom2": ["A", 1, "N"]}}]}
\end{verbatim}
\end{quote}
The two design intervals cover the terminal edit core exactly, and the bond
constraint declares the head-to-tail amide bond between \texttt{A22:C} and
\texttt{A1:N}---the bond channel this task type exists to test. The returned
candidate closes the ring: exactly one new bond is formed and
$d_{\mathrm{N\text{-}C}} = 1.32$\,\AA{}, inside the $[1.20, 1.50]$\,\AA{}
band, so the target gate passes (Figure~\ref{fig:cyclization-example}), and
both substitutions stay inside the core. The scope gate nevertheless fails
the candidate: ring closure relaxes the protected backbone by up to
$0.21$\,\AA{} against the $0.10$\,\AA{} tolerance. This failure mode is
mechanical rather than a compilation error---indeed, no candidate from any
backend passes all three gates on this task type in the frozen run, because
a genuinely closed ring cannot leave fourteen protected residues rigid.

\subsection{Generated structures}
\label{app:figures}

\begin{figure}[t]
\centering
\includegraphics[width=0.8\linewidth]{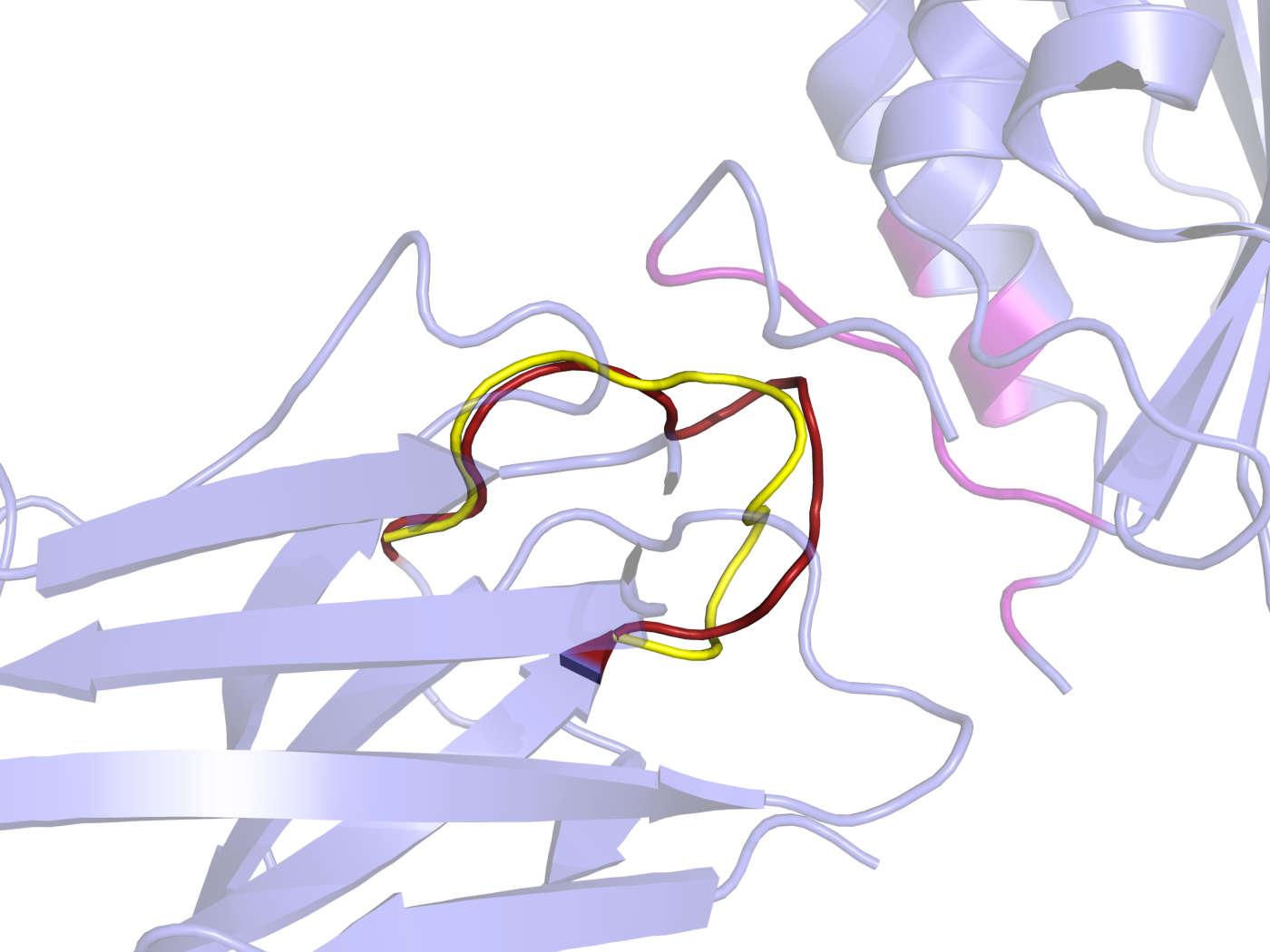}
\caption{Inter-domain approach, successful.
The source and candidate structures are overlaid in a single view.
The editable region (red, \texttt{B:99-110})
moves toward the reference patch (purple, on chain A). The dashed line marks
the closest patch contact of edit-core residue \texttt{B:101}
(to \texttt{A:201}), which shortens from $10.7$\,\AA{} to $6.0$\,\AA{}, and
the median source-referenced distance drops from $8.95$\,\AA{} to
$7.48$\,\AA{} ($\Delta = -1.43$\,\AA{}, gate requires $\le -1.0$\,\AA{}).
All three gates pass.}
\label{fig:move-example}
\end{figure}

\begin{figure}[t]
\centering
\includegraphics[width=0.49\linewidth]{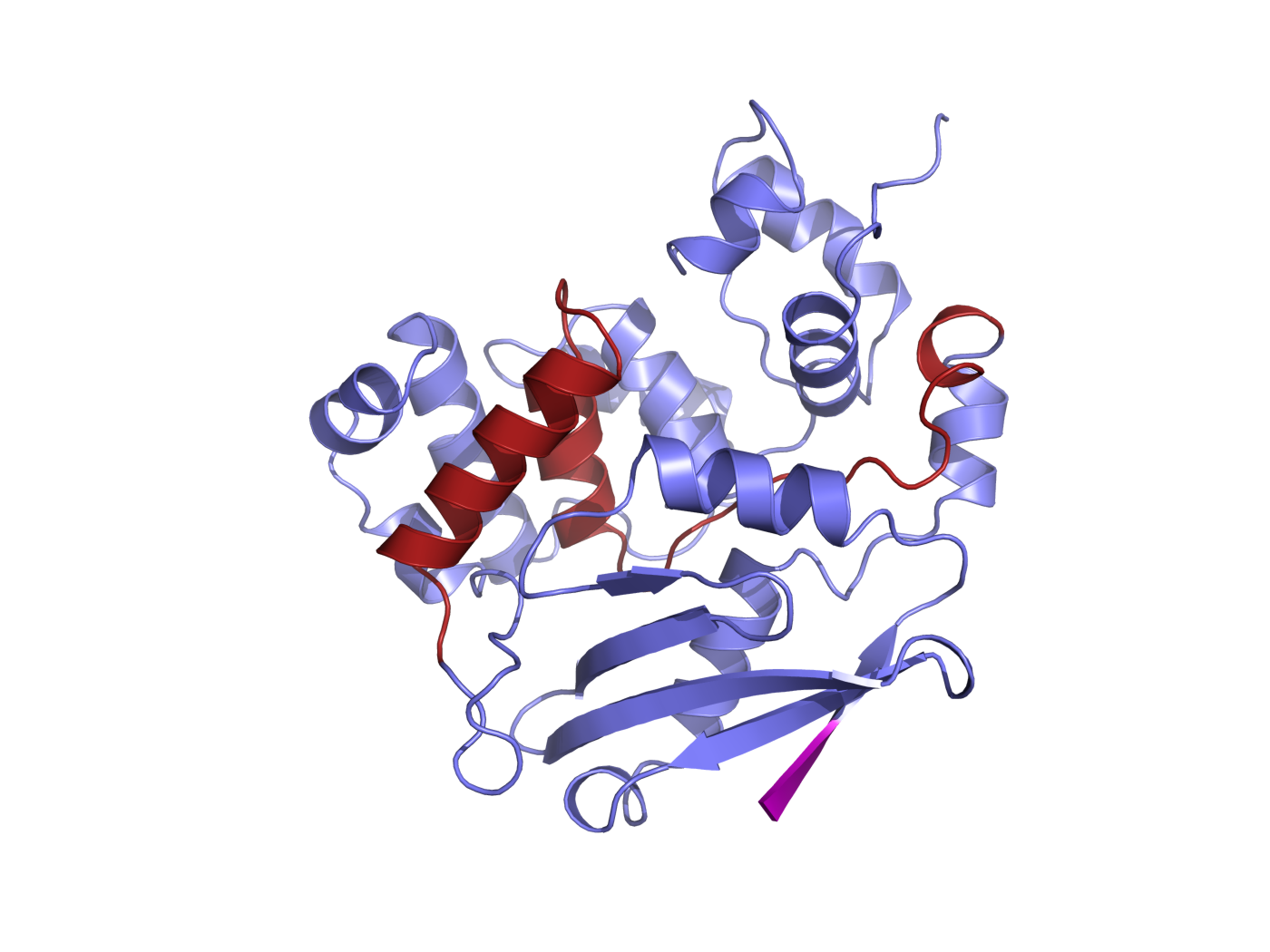}\hfill
\includegraphics[width=0.49\linewidth]{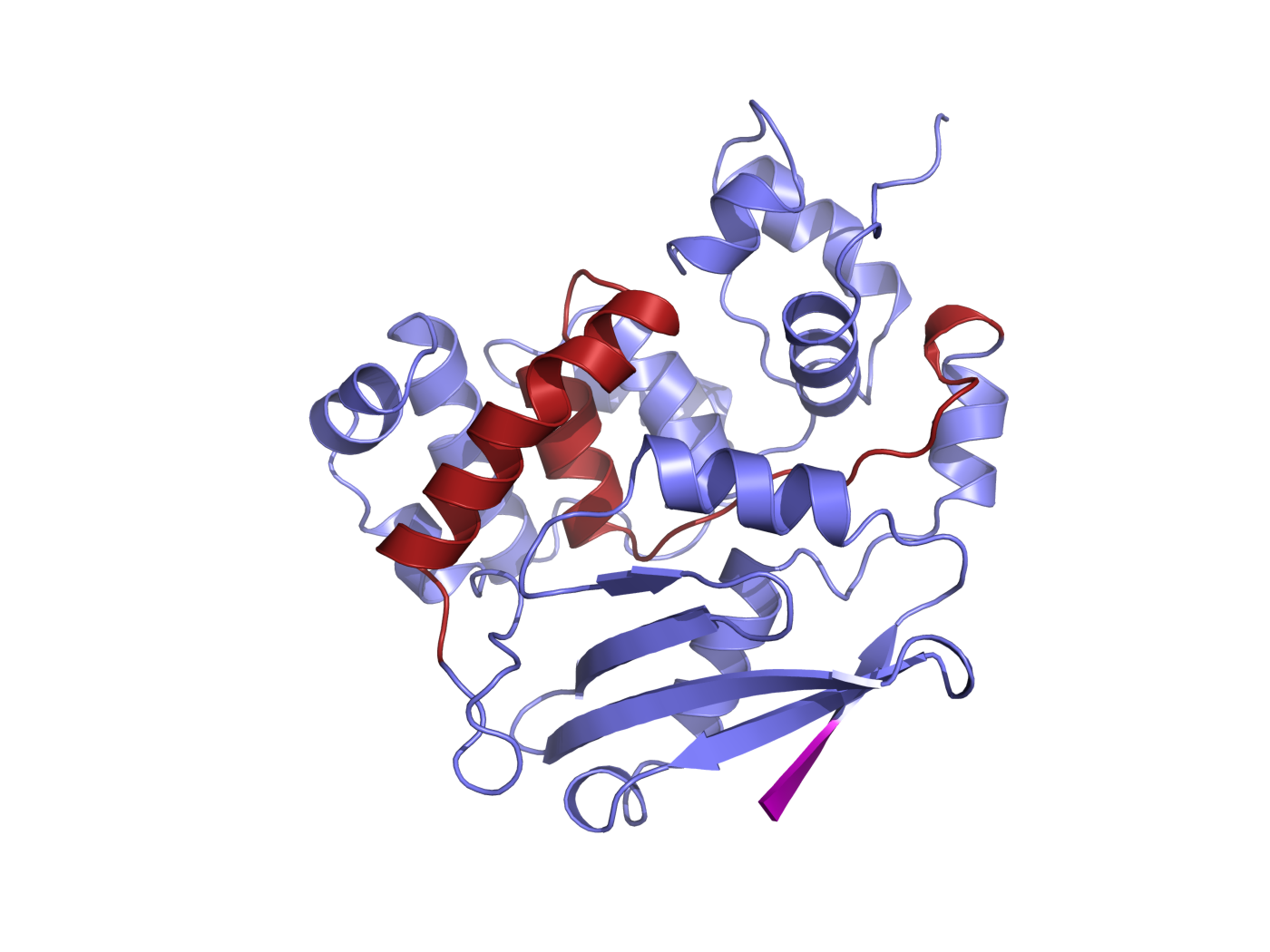}
\caption{Helix enrichment, successful. The
$48$-residue editable region (red, \texttt{A:97-144}) is assigned no helix
in the source (DSSP helix fraction $0.00$ over the window); in the candidate
it is remodelled into helices covering $56\%$ of the window (helix fraction
$0.56$, balance $0.54 \ge 0.35$), with $42$ of $48$ residues substituted
inside the core and the protected context emitted unchanged (maximum
protected-backbone displacement $3\times10^{-5}$\,\AA{}). All three gates
pass.}
\label{fig:alpha-example}
\end{figure}

\begin{figure}[t]
\centering
\includegraphics[width=0.49\linewidth]{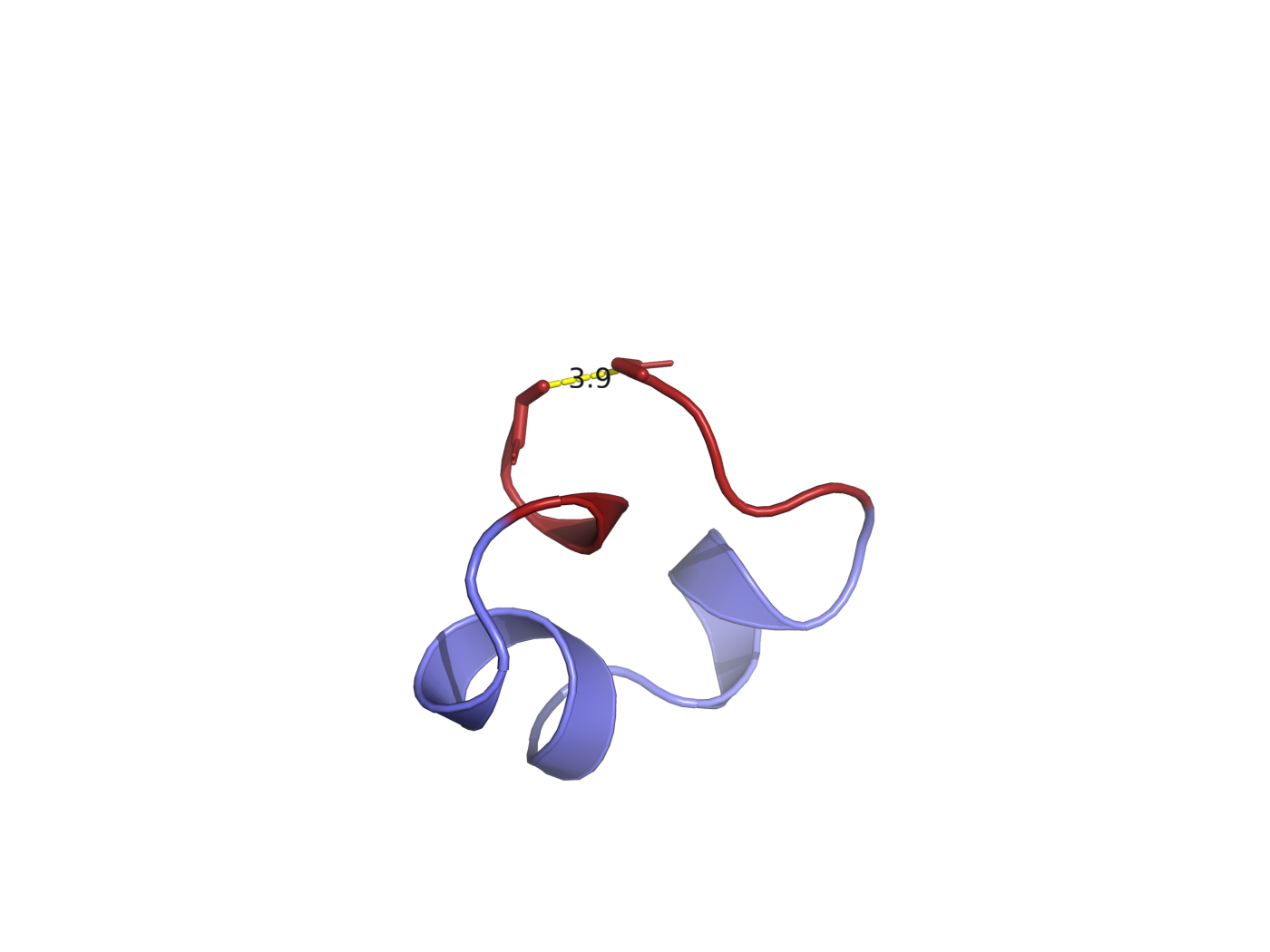}\hfill
\includegraphics[width=0.49\linewidth]{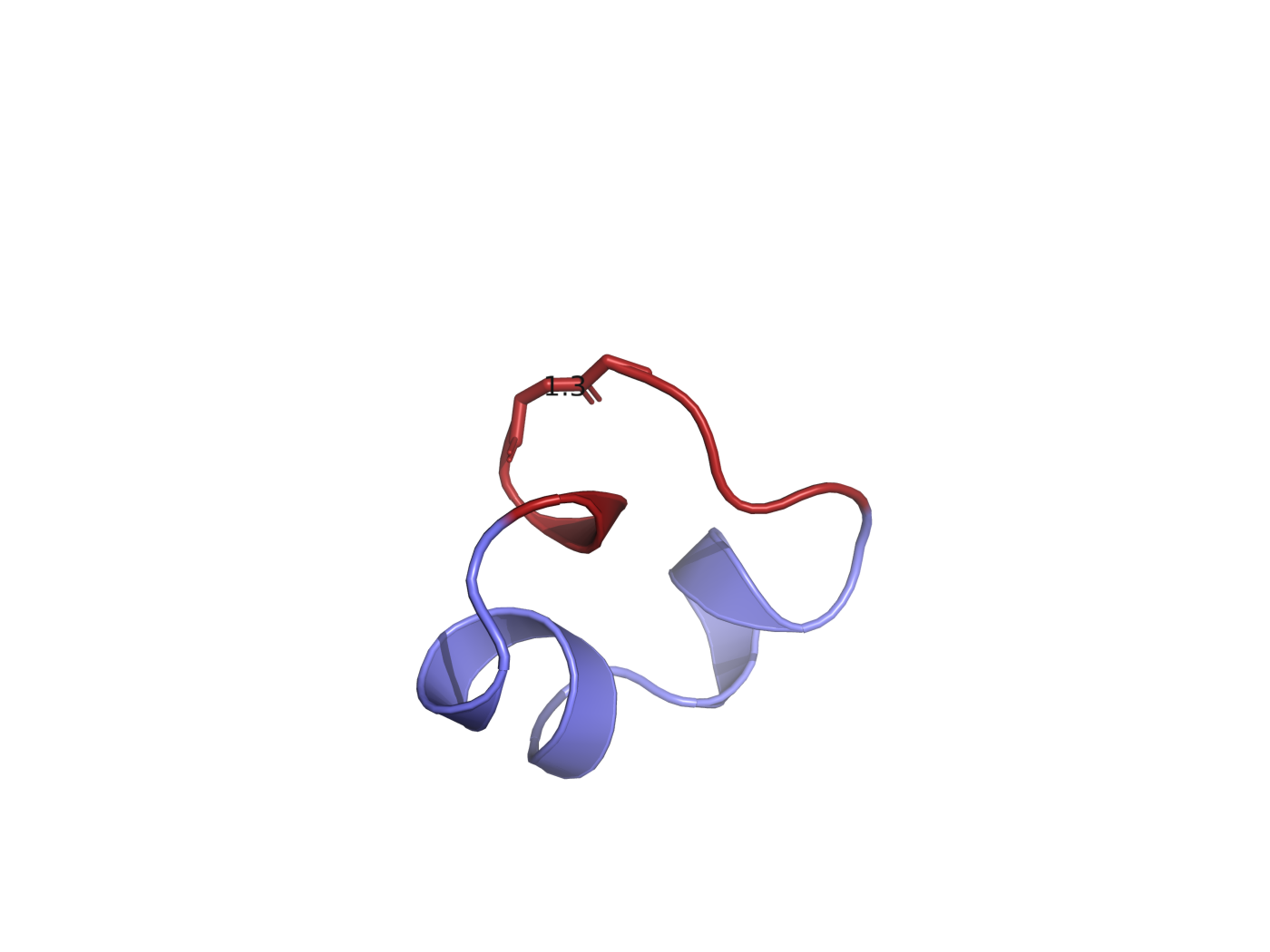}
\caption{Head-to-tail cyclization. The
$22$-residue peptide's edit core (red) spans the first and last four
residues; the terminal residues are drawn as sticks and the dashed line
marks the N$\cdots$C distance. Left: opened source ($3.9$\,\AA{}); right:
candidate, in which the declared head-to-tail amide bond closes the ring
($d_{\mathrm{N\text{-}C}} = 1.32$\,\AA{}, inside the $[1.20,
1.50]$\,\AA{} target band; exactly one new bond). The target gate passes;
the scope gate does not: closing the ring relaxes the protected backbone by
up to $0.21$\,\AA{} (tolerance $0.10$\,\AA{}), while both substitutions
stay inside the core and none of the four structural invariants changes.}
\label{fig:cyclization-example}
\end{figure}

Figures~\ref{fig:move-example}--\ref{fig:cyclization-example} render the
source and candidate structures of the three rollouts above (PyMOL cartoons;
editable region in firebrick red, reference patch in purple, remaining
protected context in slate; the cyclization example additionally draws the
two terminal residues as sticks and marks the N$\cdots$C distance).
Candidates are aligned to the source by a rigid fit over protected
C$\alpha$ anchors (post-alignment anchor RMSD $0.000$\,\AA{} for the two
redesign examples; $0.075$\,\AA{} for the cyclization example, whose ring
closure necessarily relaxes the protected backbone).

\subsection{Cyclization decomposition: shared scope candidates and closure}
\label{app:cyclization-decomposition}

The head-to-tail family contains 50 tasks and is expressible only on a backend
with an explicit covalent-bond primitive.  We therefore report the BoltzGen
lane separately and mark ODesign and RFd3 as ``--'' rather than treating an
unavailable input language as a model failure.  The audit below uses the five
available model--BoltzGen score files; the Gemini score file was not present at
the time of this appendix freeze.  Each model contributes 850 auditable
candidate rows (17 candidates per task).  ``Topology'' requires exactly one
new native $A{:}22{:}C$--$A{:}1{:}N$ head-to-tail bond; ``geometry'' additionally
requires a $1.20$--$1.50$\,\AA{} terminal distance; ``scope'' is the protected
context gate; and ``joint'' is the same-candidate conjunction.

\begin{table*}[t]
\centering
\small
\setlength{\tabcolsep}{4.4pt}
\renewcommand{\arraystretch}{1.12}
\begin{tabular}{@{}lrrrrrr@{}}
\toprule
\textbf{Model} & \textbf{Topology} & \textbf{Geometry} & \textbf{Scope} &
\textbf{Joint} & \textbf{Any geom.} & \textbf{Any scope} \\
 & \multicolumn{1}{c}{850/850} & \multicolumn{1}{c}{/850} & \multicolumn{1}{c}{/850} &
\multicolumn{1}{c}{tasks / 50} & \multicolumn{1}{c}{tasks / 50} & \multicolumn{1}{c}{tasks / 50} \\
\midrule
Claude Opus 5   & 100.0\% & 774 (91.1\%) & 50 (5.9\%) & 0 & 50 & 50 \\
DeepSeek V4 Pro & 100.0\% & 773 (90.9\%) & 50 (5.9\%) & 0 & 50 & 50 \\
GPT-5.6-Sol     & 100.0\% & 778 (91.5\%) & 50 (5.9\%) & 0 & 50 & 50 \\
Kimi K3         & 100.0\% & 766 (90.1\%) & 50 (5.9\%) & 0 & 50 & 50 \\
Qwen3.8Max     & 100.0\% & 779 (91.6\%) & 50 (5.9\%) & 0 & 50 & 50 \\
\midrule
\multicolumn{7}{@{}l}{\textit{Unsupported backends: ODesign and RFd3 are ``--'' (no expressible cyclization specification).}}\\
\bottomrule
\end{tabular}
\caption{Cyclization gate decomposition for the five available BoltzGen score
files.  Topology, geometry, and scope are candidate-level counts; ``any''
columns are task-level best-of-17 diagnostics.  ODesign and RFd3 do not expose
an explicit covalent-bond primitive and are therefore shown as unsupported
(``--'').}
\label{tab:app-cyclization-decomposition}
\end{table*}

The 50 scope-pass rows are effectively shared across models.  They are all
rollout~1; after canonicalizing atom coordinates and topology, all 50
task-matched candidates are identical across the five models.  Forty-nine of
50 CIF files are byte-identical as well.  The only byte-level difference is a
single GPT file whose occupancy metadata differs; its atom coordinates,
residue identities, and bond record are unchanged.  These rows therefore
represent the common source/adapter-preserving candidate rather than five
independent model-specific solutions.  They should not be used to claim that
one language model preserves the protected region better than another.

The model-dependent signal appears in the remaining candidates.  Qwen3.8Max
has the highest geometry-pass rate ($779/850=91.6\%$) and the smallest
non-closing tail; GPT-5.6-Sol follows at $778/850=91.5\%$.  Claude Opus 5 and
DeepSeek V4 Pro are close behind, whereas Kimi K3 has the lowest geometry-pass
rate ($766/850=90.1\%$) and the largest non-closing tail.  These differences
are small (1.5 percentage points) and arise only after the explicit bond
constraint is compiled successfully; they are not differences in protected
scope preservation.

For the closed candidates, the aligned protected-backbone displacement has a
median of approximately $0.43$--$0.45$\,\AA{}, while the frozen scope tolerance
is $0.10$\,\AA{}.  These candidates preserve length, residue order, backbone
mapping, and edit all substitutions inside the declared core, but they trigger
\texttt{protected\_coordinate\_changed}.  The unchanged scope-pass candidate
leaves the termini approximately $3.0$--$4.0$\,\AA{} apart and consequently
misses the cyclization distance.  The strict $0/50$ result is therefore the
empty intersection of two candidate sets: one set preserves the context and
one set closes the bond.

This case is a coupled-contract diagnostic.  The models can emit the required
topology and usually attain the local bond geometry, while the frozen contract
does not permit the local relaxation needed to close the termini.  A
cyclization-specific report should therefore list topology, geometry, scope,
and strict joint gates separately, and should protect only genuinely fixed
context while allowing a calibrated terminal/relax-shell motion.  The existing
CIFs can be re-scored under that relaxed cyclization contract without repeating
the GPU generation; the strict frozen-context score can remain as a secondary
stress test.

\subsection{Cyclization failure cases and accounting of false failures}
\label{app:cyclization-failure-cases}

\begin{figure*}[t]
\centering
\includegraphics[width=0.96\linewidth]{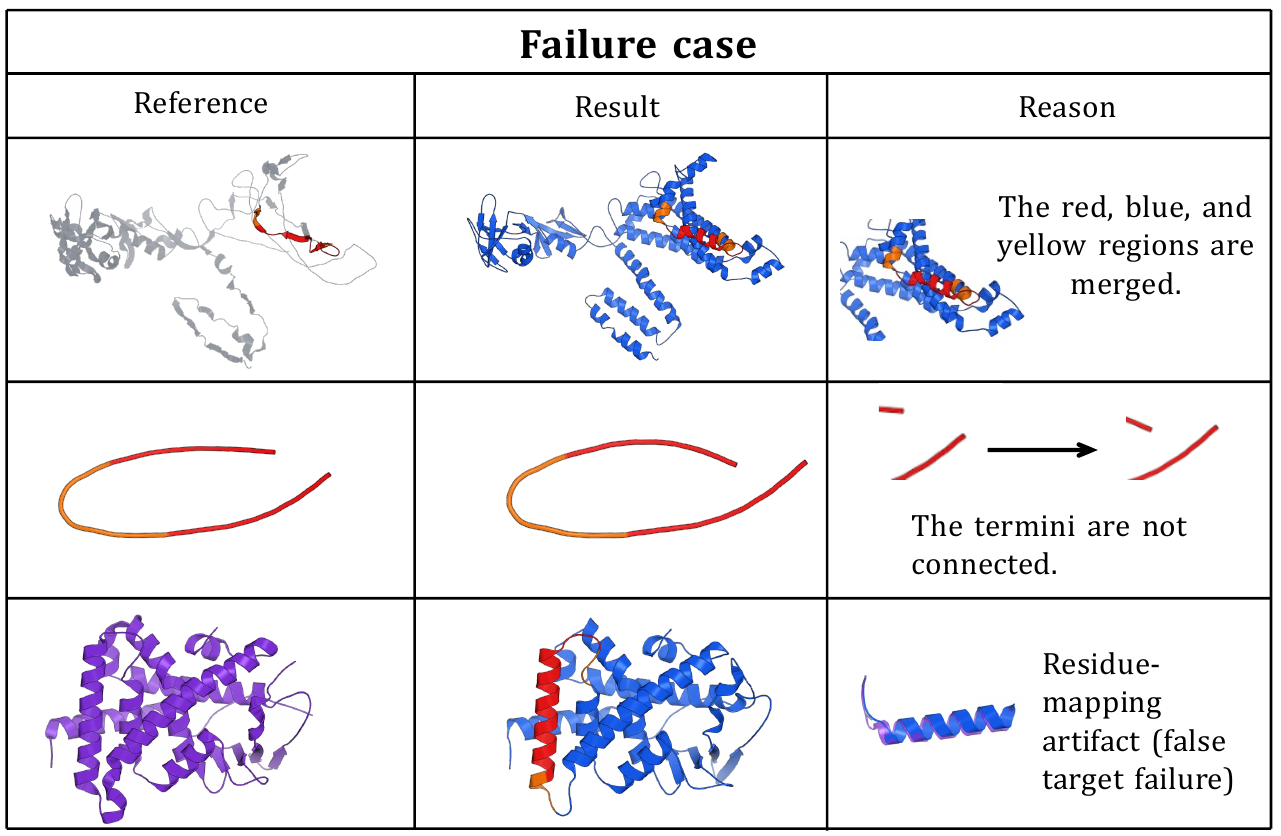}
\caption{Representative cyclization failure cases.  Top: the editable and
protected color regions are merged, so the region partition required by the
contract is not preserved.  Middle: the generated chain remains open and the
termini are not connected, which is a genuine topology/target failure once the
structure and residue mapping are valid.  Bottom: a residue-mapping artifact
creates an apparent target failure even though the visual structure is not the
source of the discrepancy; such rows must be remapped and rescored before they
are attributed to a model.}
\label{fig:cyclization-failure-cases}
\end{figure*}

The model-specific comparison above concerns candidates that pass the topology
and geometry checks.  A separate failure-case audit is needed because a visual
failure can arise at different layers of the pipeline.  Figure~\ref{fig:cyclization-failure-cases}
shows three representative patterns: a region-partition failure in which the
red, blue, and yellow regions are merged; a genuine open-chain result in which
the termini remain disconnected; and a residue-mapping artifact that creates a
false target failure.  These patterns are not interchangeable and are assigned
to different accounting terminals.  The supplied panel is a representative
diagnostic and carries no task or model identifier; we therefore do not
attribute a pictured row to a particular model without its task-level log.

\begin{table*}[t]
\centering
\small
\setlength{\tabcolsep}{5pt}
\renewcommand{\arraystretch}{1.16}
\begin{tabular}{@{}p{0.22\linewidth}p{0.22\linewidth}p{0.46\linewidth}@{}}
\toprule
\textbf{Failure pattern} & \textbf{Primary accounting terminal} & \textbf{Interpretation and treatment} \\
\midrule
Region colors merged & Region partition / scope terminal & The editable, protected, and terminal regions cannot be separated reliably.  Do not call this a cyclization target failure; retain it as a compilation or scope-contract failure and inspect the native specification. \\
\addlinespace
Termini not connected & Target topology/geometry failure & The output is an open chain.  If valid output, structural integrity, and residue mapping pass, count it as a genuine target failure; do not merge it with unsupported-backend or scope failures. \\
\addlinespace
Residue-mapping artifact & Scorer/mapping audit & The apparent target miss is caused by comparing the wrong residue correspondence.  Quarantine the row, repair the official map, and rescore it before assigning a model failure. \\
\bottomrule
\end{tabular}
\caption{Failure-case taxonomy for cyclization.  The three rows correspond to
the panels in Figure~\ref{fig:cyclization-failure-cases}; only the middle row
is a target failure by itself.}
\label{tab:cyclization-failure-taxonomy}
\end{table*}

The middle panel is the failure mode most directly relevant to the requested
cyclization primitive: the candidate does not realize a head-to-tail bond and
therefore cannot satisfy the $d_{\mathrm{N-C}}$ target.  It should be reported
separately from the closed-but-scope-violating candidates in the decomposition
table.  The top panel indicates that the agent or adapter did not preserve the
contract's region partition, while the bottom panel demonstrates why mapping
validation must precede model attribution.  In particular, a strict success
rate should not be lowered by a row that is later shown to be a scorer artifact.

Across the five available BoltzGen models, there is no non-empty set of strict
cyclization-passing samples to compare: every model has 0/50 task-level joint
success.  The model-specific differences are instead in the geometry-pass
candidate sets (Qwen3.8Max $779/850$, GPT-5.6-Sol $778/850$, Claude Opus 5
$774/850$, DeepSeek V4 Pro $773/850$, and Kimi K3 $766/850$).  The 50
scope-pass candidates are shared after canonicalization (49/50 CIFs are also
byte-identical), so they provide no evidence for a model-specific protected
context advantage.  Gemini is not included in this comparison because its
BoltzGen score file was not available at the appendix freeze.

For the failure-case figure, the correct conclusion is therefore layered:
models almost always compile the explicit bond primitive and usually produce
local closure geometry, but strict cyclization remains blocked either by an
open-chain target miss, by a region/scope violation, or by an evaluation-map
artifact.  These terminals should remain separate in the final score table.

\subsection{Gate evaluation, in pseudocode}
\label{app:gatepseudo}

Algorithm~\ref{alg:gates} gives the evaluator at the level of the frozen
scorer. Algorithm~\ref{alg:target-detail} spells out the per-edit-type
target criteria behind GateTarget, and Algorithm~\ref{alg:motif-gates}
records the three-gate mirror applied to the imported motif-scaffolding
cases. All thresholds are the frozen profile values; card-carried overrides
from earlier profiles are ignored.

\begin{algorithm}[t]
\caption{Three-gate evaluation of one candidate}
\label{alg:gates}
\begin{algorithmic}[1]
\Require task $\tau=(s,X,\mathcal{M},u)$, source rows, candidate rows
\Function{GateTarget}{$\tau$, src, cand}
  \If{$\tau.\mathrm{action}$ = \emph{approach / separation}}
          \Comment{per-type criteria: Algorithm~\ref{alg:target-detail}}
    \State $A \gets$ anchors $=$ protected C$\alpha$ away from core and patch
    \State $T \gets \mathrm{KabschFit}(\mathrm{cand}|_A \to \mathrm{src}|_A)$
    \State $\hat D \gets \mathrm{median}_{i \in \mathcal{M}_{\mathrm{edit}}}
            \min_{j \in \mathcal{M}_{\mathrm{ref}}} \lVert T\hat c_i - c_j\rVert$
            \Comment{patch points from the source}
    \State $D \gets$ same on the source core
    \State \Return ($\hat D - D \le -1.0$) if approach else ($\hat D - D \ge +1.0$)
  \ElsIf{$\tau.\mathrm{action}$ = \emph{helix / sheet}}
    \State $(f_H, f_E) \gets$ DSSP fractions over cand's editable region
    \State $b \gets f_H - f_E$ if helix else $f_E - f_H$
    \State \Return $b \ge 0.35$ (helix) \;/\; $b \ge 0.10$ (sheet)
            \Comment{absolute level of the candidate}
  \ElsIf{$\tau.\mathrm{action}$ = \emph{mimicry}}
    \State \Return $\mathrm{KabschRMSD}(\text{mapped N/CA/C/O}) \le 2.0$
            $\wedge$ mapped $\ge 3$ $\wedge$ coverage $\ge 0.80$
  \ElsIf{$\tau.\mathrm{action}$ = \emph{burial}}
    \State $r \gets$ per-residue RSA (Shrake--Rupley, Tien) over cand
    \State \Return $\mathrm{median}(r|_{\mathrm{core}}) - \mathrm{median}(r^{\mathrm{src}}|_{\mathrm{core}}) \le -0.10$
            $\wedge$ $\mathrm{median}(r|_{\mathrm{core}}) \le 0.40$
            $\wedge$ $R_g$ ratio $\in [0.70, 0.95]$
  \Else \Comment{cyclization}
    \State \Return exactly one new head-to-tail amide bond
            $\wedge$ $d_{\mathrm{N-C}} \in [1.20, 1.50]$
            $\wedge$ topology evidence complete
  \EndIf
\EndFunction
\Statex
\Function{GateRefold}{$\tau$, cand}
  \If{$\tau.\mathrm{action} =$ cyclization} \Return \textsc{n/a} \EndIf
  \State motif $\gets$ protected residues of the edit chain, frozen
  \State $S \gets$ ProteinMPNN redesign of the editable region ($k{=}4$)
  \For{$s \in S$} \Comment{ESMFold refold of each sequence}
    \If{$\mathrm{RMSD}_{\mathrm{chain}}(s) < 3.0$ \textbf{and}
        $\mathrm{RMSD}_{\mathrm{motif}}(s) < 2.0$} \Return \textsc{pass} \EndIf
  \EndFor
  \State \Return \textsc{fail}\Comment{no passing sample found in budget}
\EndFunction
\Statex
\Function{GateScope}{$\tau$, src, cand}
  \State \Return length, chains, order, N/CA/C/O mapping unchanged
          $\wedge$ identity outside $\mathcal{M}_{\mathrm{edit}}$ unchanged
          (three-letter comparison) $\wedge$ canonical residues only
          $\wedge$ protected backbone $\le 0.1$\,\AA{} after anchored fit
\EndFunction
\Statex
\State \textbf{strict} $\gets$ GateTarget $\wedge$ GateRefold $\wedge$ GateScope
\State terminals (reported separately, never folded into zeros):
       unsupported-by-backend $>$ backend-failure $>$ invalid-spec $>$
       gate failures; missing refold evidence is \emph{pending}, not failure
\end{algorithmic}
\end{algorithm}

\begin{algorithm}[t]
\caption{GateTarget per edit type (the switch of
Algorithm~\ref{alg:gates} expanded; frozen profile values)}
\label{alg:target-detail}
\begin{algorithmic}[1]
\Require candidate aligned to the source by the anchored fit $T$ of
         Algorithm~\ref{alg:gates}; masks
         $\mathcal{M}_{\mathrm{edit}}$, $\mathcal{M}_{\mathrm{ref}}$; $D$
         = the source value of the case metric
\If{$\tau.\mathrm{action}$ = \emph{inter-domain approach / separation}}
  \State metric: median over edit-core residues of the nearest pseudo-CB
         distance to reference-patch points
         ($\hat D$, computed as in Algorithm~\ref{alg:gates})
  \State \Return $\hat D - D \le -1.00$\,\AA{} (approach)\;\;/\;\;
         $\hat D - D \ge +1.00$\,\AA{} (separation)
\ElsIf{$\tau.\mathrm{action}$ = \emph{helix / sheet enrichment}}
  \State $(f_H, f_E) \gets$ DSSP residue fractions over the candidate's
         editable region (mkdssp when available; the deterministic
         $\phi/\psi$ heuristic serves only as per-file fallback)
  \State \Return $f_H - f_E \ge 0.35$ (helix)\;\;/\;\;
         $f_E - f_H \ge 0.10$ (sheet)
\ElsIf{$\tau.\mathrm{action}$ = \emph{reference-fragment mimicry}}
  \State $R \gets$ backbone Kabsch RMSD of the mapped edit-core residues
         (N/CA/C/O) against the reference fragment
  \State \Return $R \le 2.00$\,\AA{} $\wedge$ mapped $\ge 3$ $\wedge$
         coverage of the editable region $\ge 0.80$
\ElsIf{$\tau.\mathrm{action}$ = \emph{shape compaction (v3)}}
  \State $\delta_{\mathrm{SASA}} \gets 1 -
         \sum_{i \in \mathcal{M}_{\mathrm{edit}}} \mathrm{SASA}^{\mathrm{cand}}_i \,/\,
         \sum_{i \in \mathcal{M}_{\mathrm{edit}}} \mathrm{SASA}^{\mathrm{src}}_i$
         \Comment{like-for-like atoms: backbone$+$CB when either side lacks side chains}
  \State \Return $\delta_{\mathrm{SASA}} \ge 0.10$ $\wedge$
         $R_g^{\mathrm{cand}}/R_g^{\mathrm{src}} \in [0.70, 0.95]$
         \Comment{anti-collapse guard; the v2 asphericity leg is diagnostic only}
\ElsIf{$\tau.\mathrm{action}$ = \emph{core burial (v3)}}
  \State $r \gets$ per-residue relative solvent accessibility over the
         candidate (Shrake--Rupley, Tien)
  \State \Return $\mathrm{median}(r|_{\mathcal{M}_{\mathrm{edit}}}) -
         \mathrm{median}(r^{\mathrm{src}}|_{\mathcal{M}_{\mathrm{edit}}}) \le -0.10$
         $\wedge$ $\mathrm{median}(r|_{\mathcal{M}_{\mathrm{edit}}}) \le 0.40$
         $\wedge$ $R_g^{\mathrm{cand}}/R_g^{\mathrm{src}} \in [0.70, 0.95]$
\Else \Comment{head-to-tail cyclization}
  \State \Return exactly one new head-to-tail amide bond $\wedge$
         $d_{\mathrm{N-C}} \in [1.20, 1.50]$\,\AA{} $\wedge$
         adapter topology evidence complete
\EndIf
\Statex
\State no target case exists for motif scaffolding inside the agent
       lineage: the imported contract is evaluated by
       Algorithm~\ref{alg:motif-gates}
\end{algorithmic}
\end{algorithm}

\begin{algorithm}[t]
\caption{Motif-scaffolding evaluation (imported MotifBench output
contract; the three-gate mirror of Algorithm~\ref{alg:gates})}
\label{alg:motif-gates}
\begin{algorithmic}[1]
\State \textbf{Gate 1, assembly:} format valid (single-chain mmCIF, exactly
       one model) $\wedge$ placement valid (declared total length and segment
       order; scaffold gaps are the model's own design, audited only for
       arithmetic self-consistency with the public total length)
\State \textbf{Gate 2, sidecar refold:} any of 8 ProteinMPNN
       $\rightarrow$ ESMFold designed sequences refolds with whole-chain
       backbone RMSD $< 2.0$\,\AA{} and motif RMSD $< 1.0$\,\AA{}
\State \textbf{Gate 3, motif preservation:} motif identity valid $\wedge$
       per-segment motif backbone RMSD $\le 2.0$\,\AA{} against the source
       motif (Kabsch over N/CA/C/O) $\wedge$ sequence changes confined to
       scaffold positions
\State \Return success $=$ Gate 1 $\wedge$ Gate 2 $\wedge$ Gate 3;
       missing sidecar evidence never passes and is never imputed
\end{algorithmic}
\end{algorithm}

\paragraph{Open-design run (status at freeze).}
The open-scaffold caliber above---canonical gaps withheld from the
instruction, the declared placement audited only for arithmetic
self-consistency---was evaluated with the same six language models and
three backends as Section~\ref{sec:experiments}, on all 30 imported cases
per configuration. In the frozen 18-configuration run, grounding, the
declared-placement assembly gate, and the motif-preservation gate saturate
at $30/30$ everywhere: the two deterministic gates no longer separate
models on the imported family, and the discriminative burden rests entirely
on the sidecar refold gate, whose scoring was pending at the freeze.
Consistent with Algorithm~\ref{alg:motif-gates}, no motif-scaffolding task
is counted as a success in the absence of sidecar evidence. Two
configuration caveats apply: the RFdiffusion3 adapter compiles with fully
fixed motif sequences (an upstream defect), so its redesign capability is
not exercised; and Claude Opus~5 requests omit the temperature parameter,
which its provider rejects. The one-sentence instruction quoted in
Table~\ref{tab:card-instances} is the condensed deployment of the same
contract; under it, grounding completes on $178/180$ model--task pairs,
with each of two models failing a single task.

\subsection{Qualitative Case Studies: Spec-Space Planning Disparity}
\label{app:case-studies}

To understand how high-level biological reasoning translates into low-level backend parameterizations, Figure~\ref{fig:case-study} illustrates two representative compilation episodes executed on the ODesign backend. Both cases expose how subtle differences in the compiled specification reflect fundamentally different levels of protein understanding, directly determining macroscopic task outcomes.

\begin{figure*}[t]
\centering
\small

\noindent
\begin{minipage}{\linewidth}
\subsubsection*{(a) Case 1 --- Separation: Hotspot Curation}
\vspace{2pt}
\noindent\textbf{Task.} \emph{``Move the red edit core away from the purple reference; substitutions only inside the edit core.''}\\
\noindent\textbf{Context.} Edit core: \texttt{B:67--74} ($\sim 310$\,kcal interface); reference patch: \texttt{A:\{62, 64, 95--97, 101--102, 141, 143, 146--148\}}.
\vspace{6pt}

\noindent
\begin{tabularx}{\linewidth}{@{} l >{\raggedright\arraybackslash\ttfamily\scriptsize}X @{}}
\toprule
\textbf{Model} & \textbf{Compiled Output (JSON)} \\
\midrule
Claude-Opus-5 & \{"hotspot": "A/62, A/141, A/147", "partial\_diff": "B/59--82"\} \\
\addlinespace[3pt]
Kimi-K3       & \{"hotspot": "A/62, A/64, A/95, A/96, A/97, A/101, A/102, A/141, A/143, A/146, A/147, A/148", "partial\_diff": "B/59--82"\} \\
\addlinespace[3pt]
Qwen3.8Max    & \{"hotspot": "A/62", "partial\_diff": "B/67--74"\} \\
\bottomrule
\end{tabularx}
\vspace{6pt}

\noindent\emph{Biophysical Mechanism.} ODesign's \texttt{hotspot} primitive defines the spatial conditioning vector toward which the redesigned segment is steered. Claude-Opus-5 intentionally curates \textbf{3 of the 12} candidate patch residues, selecting a coherent spatial cluster on the distal interface that creates an effective pulling vector. In contrast, Kimi-K3 verbatim transcribes \textbf{all 12} residues without geometric selection; because the landmarks envelop the interface, their isotropic conditioning vectors cancel out. Conversely, Qwen3.8Max over-prunes the specification down to an isolated \textbf{single residue}, providing an insufficient conditioning signal. While all three models satisfy the syntax format, only Claude-Opus-5 encodes an authentic \emph{directional comprehension} of interface mechanics.
\end{minipage}

\vspace{12pt}
\hrulefill
\vspace{8pt}

\noindent
\begin{minipage}{\linewidth}
\subsubsection*{(b) Case 2 --- Approach: Boundary Field Auditing}
\vspace{2pt}
\noindent\textbf{Task.} \emph{``Move the red edit core closer to the purple reference.''}\\
\noindent\textbf{Context.} Edit core: \texttt{B:99--110}; relax shell: \texttt{B:91--98, 111--118}; reference patch: \texttt{A:\{7--12, 193, 196--197, 201\}}.
\vspace{6pt}

\noindent
\begin{tabularx}{\linewidth}{@{} l >{\raggedright\arraybackslash\ttfamily\scriptsize}X @{}}
\toprule
\textbf{Model} & \textbf{Compiled Output (JSON)} \\
\midrule
Claude-Opus-5 & \{"design": "B/1--98, 12--12, B/111--128", "hotspot": "A/8..A/201 (9 res)", "partial\_diff": "B/99--110"\} \hfill \rmfamily\itshape(Core only) \\
\addlinespace[3pt]
DSV/Kimi/Qwen & \{"design": "B/1--98, 12--12, B/111--128", "hotspot": "A/7..A/201 (9 res)", "partial\_diff": "B/91--118"\} \hfill \rmfamily\itshape(Core + shell) \\
\bottomrule
\end{tabularx}
\vspace{6pt}

\noindent\emph{Biophysical Mechanism.} All four models construct structurally viable backbone design spans and correct hotspot selections. The sole divergence lies in the \texttt{partial\_diff} field, which defines the continuous coordinates subject to generative denoising. Claude-Opus-5 confines diffusion strictly to the editable core (\texttt{B:99--110}). In contrast, DeepSeek-V4-Pro, Kimi-K3, and Qwen3.8Max conflate the editable core with the surrounding \emph{relaxation shell}, expanding diffusion across \texttt{B:91--118}. Diffusing through protected residues inevitably alters invariant coordinates and triggers sequence redesign outside the legal window (\texttt{substitution\_outside\_edit\_core}). Consequently, \textbf{Claude-Opus-5 uniquely solves the task}, whereas the other models fail the $G_3$ scope contract completely.
\end{minipage}

\vspace{8pt}
\caption{Two real compilation episodes on the ODesign backend illustrating how biomolecular spatial reasoning manifests in specification space. \textbf{Top (Case 1):} Hotspot curation reflects directional planning across complex interfaces, where uncurated landmarks cancel out directional guidance. \textbf{Bottom (Case 2):} Boundary auditing serves as the sole discriminating factor, where confusing the editable core with the relaxation shell results in catastrophic $G_3$ scope failure.}
\label{fig:case-study}
\end{figure*}

\subsection{Execution Bottlenecks and Biophysical Attrition Analysis}
\label{app:bottlenecks}



\paragraph{Target Attainment vs. Off-Target Scope Leakage.}
Crucially, reaching the assigned target metric ($G_1$) does not constitute a valid
edit if the surrounding scaffold is deformed. This distinction is most dramatic in
RFdiffusion3: while the model successfully drives the editable core toward the
requested geometry in 96.0 tasks (27.4\%), it maintains scope fidelity in only 51.0
tasks (14.6\%). As a direct result, joint geometric compliance collapses to just
26.0 tasks (\textbf{7.4\%}). Unconstrained diffusion backbones routinely achieve
local structural objectives by dispersing structural strain into invariant motifs,
proving that strict coordinate-level boundary auditing ($C_\alpha\text{--RMSD} \le 0.5\,\text{\AA}$)
is indispensable to prevent global conformational drift. Conversely, BoltzGen and
ODesign provide substantially tighter contextual conditioning, retaining 94.5\% and
80.6\% scope fidelity, respectively.
\subsection{Special Setting in motif scaffolding} \label{app:motif2}
\paragraph{Budget Scaling Dynamics ($s@1 \to s@16$) and Kinetic Traps.}
Tracking the scaling trajectory from $s@1$ to $s@16$ underscores the role of structural diversity in overcoming conformational energy barriers. 
Across all models, scaling the candidate budget yields substantial performance amplification (e.g., a $>3\times$ gain for Claude-Opus-5). 
Scaffolding isolated motifs requires navigating highly rugged, non-convex thermodynamic landscapes, where single-candidate sampling ($s@1$) frequently falls into kinetically trapped local minima. 
Expanding the budget supplies the conformational entropy necessary to sample viable alternative topologies, markedly boosting the recovery of physically self-consistent carriers.

\paragraph{Diagnostic Bottlenecks and the Sim-to-Real Horizon.}
Fine-grained error diagnostics across contextual editing (Table~\ref{tab:family-main}) and de novo scaffolding (Table~\ref{tab:motif-bounds}) reveal that failure modes stem not from high-level intent parsing, but from coupled spatial planning (58.7--61.6\% failure) and the co-optimization of local deformation with contextual preservation (73.0--82.1\% failure; see Appendix~\ref{app:bottlenecks} for per-tier failure taxonomies). 
While baseline agents (such as Kimi-K3) stumble early on specification compilation, frontier models (Claude-Opus-5 and GPT-5.6-Sol) predominantly fail by over-deforming conserved residues to fulfill target geometries. 
Crucially, syntactic validity and isolated metric attainment repeatedly conceal these physical defects. 
Together, while our findings establish that frontier LLMs possess genuine 3D spatial planning competencies, translating these blueprints into physical molecules remains bottlenecked by generator scope leakage and thermodynamic refolding barriers (detailed downstream biophysical analyses in Appendix~\ref{app:bottlenecks}). 
Bridging this simulation-to-reality gap via physics-informed sequence--structure co-design defines the primary horizon for autonomous biomolecular agents.

\end{document}